\documentclass[sigconf]{acmart}

\AtBeginDocument{
  }

\makeatletter
\let\@authorsaddresses\@empty
\makeatother                                     
\renewcommand\footnotetextcopyrightpermission[1]{%
  \footnotetext[0]{This paper is accepted at the ACM International Conference on Supercomputing (ICS 2026), July 06--09, 2026, Belfast, United Kingdom}%
}
\copyrightyear{2026}
\acmYear{2026}
\setcopyright{cc}
\setcctype{by}
\acmConference[ICS '26]{2026 International Conference on Supercomputing}{July 06--09, 2026}{Belfast, United Kingdom}
\acmBooktitle{2026 International Conference on Supercomputing (ICS '26), July 06--09, 2026, Belfast, United Kingdom}
\acmDOI{10.1145/3797905.3807862}
\acmISBN{979-8-4007-2522-7/2026/07}

\usepackage{algorithm}
\usepackage{algorithmic}
\usepackage{amsmath}

\usepackage{amssymb}
\usepackage{booktabs}
\usepackage{multirow}
\usepackage{graphicx}
\usepackage{subcaption}
\usepackage{xcolor}
\usepackage{dblfloatfix}
\usepackage{graphicx}
\usepackage{xspace}
\usepackage{booktabs}
\usepackage{makecell}

\newenvironment{observationbox}{%
  \par\vspace{12pt}\noindent\begin{tabular}{|p{0.95\columnwidth}|}\hline\rule{0pt}{2.5ex}%
}{%
  \\[0.5ex]\hline\end{tabular}\par\vspace{12pt}\par%
}

\usepackage{listings}
\usepackage{xcolor}

\newcommand{\req}[1]{\fbox{\bfseries R#1}}

\begin{document}

\title[TuniQ: Autotuning Compilation Passes for Quantum Workloads at Scale for Effectiveness and Efficiency]{TuniQ: Autotuning Compilation Passes for Quantum \\Workloads at Scale for Effectiveness and Efficiency}

\author{Mohammad Abrarul Hasanat}
\orcid{0009-0000-6112-6107}
\affiliation{%
  \institution{University of Utah}
  \city{Salt Lake City}
  \state{Utah}
  \country{USA}
}
\email{u1592727@utah.edu}

\author{Jason Ludmir}
\orcid{0000-0002-3703-3282}
\affiliation{%
  \institution{Rice University}
  \city{Houston}
  \state{Texas}
  \country{USA}
}
\email{jzl2@rice.edu}

\author{Tirthak Patel}
\orcid{0000-0003-3127-5931}
\affiliation{%
  \institution{Rice University}
  \city{Houston}
  \state{Texas}
  \country{USA}
}
\email{tp53@rice.edu}

\author{Rohan Basu Roy}
\orcid{0000-0002-1082-9846}
\affiliation{%
  \institution{University of Utah}
  \city{Salt Lake City}
  \state{Utah}
  \country{USA}
}
\email{rohanbasuroy@sci.utah.edu}
\begin{abstract}

\textit{Quantum processors are being integrated into HPC ecosystems as co-processors, where compilation of quantum circuits into hardware-executable form determines both output fidelity and runtime. Current compilers use a fixed pass sequence and ignore the fact that optimal pass selection varies with circuit, hardware, and noise conditions. We present TuniQ, a reinforcement learning-based system that selects compilation passes at each pipeline stage, adapting to circuit, backend, and current noise profile. TuniQ introduces several novel design components like a dual-encoder for stage-aware representation, shaped rewards for cross-stage credit assignment, and dynamic action masking for valid compilation. Evaluated across diverse quantum workloads on multiple IBM Quantum Cloud processors, TuniQ improves fidelity and reduces compilation time over the state-of-the-art IBM Qiskit transpiler, generalizes across backends without retraining, and scales strongly to utility-scale circuits with growing advantage.}

\end{abstract}


\maketitle

\newcommand{\methodname}{TuniQ\xspace}
\newcommand{\sol}{TuniQ\xspace}
\newcommand{\qfast}{\textsc{Qiskit Time Optimized}\xspace}

\newcommand{\qbest}{Qiskit Fidelity Optimized\xspace} 
\newcommand{\ESP}{Expected Success Probability\xspace}

\newcommand{\tvd}{Total Variation Distance\xspace}

\vspace{-2mm}
\section{Introduction}
\label{sec:intro}

\noindent \textbf{Quantum compilation in HPC-quantum workflows.}
Quantum processors are being integrated into HPC ecosystems as co-processors~\cite{beck2024integrating, giortamis2025qonductor, alexeev2024quantum, chundury2025scaling, mahesh2025conqure,
xu2025gpu}, where quantum circuits function as kernels dispatched from classical nodes. These quantum programs, expressed as circuits, cannot execute directly on hardware: different devices support different native gate sets, and multi-qubit operations are restricted to physically adjacent qubits~\cite{li2019tackling, murali2019noise}. A compilation step, called \textit{transpilation}, transforms a logical circuit into a hardware-executable physical circuit and runs on classical HPC resources such as GPU clusters~\cite{xu2024atlas,fu2024surpassing} and multi-core nodes.

Beyond satisfying hardware constraints, compilation determines the \textit{runtime and depth} of quantum execution and which physical qubits execute each operation~\cite{wilson2020just}. Runtime is a first-order concern for near-term devices and also for fault-tolerant quantum computing (FTQC). Quantum error correction (QEC) incurs substantial overhead, and early FTQC systems will rely on shallow logical circuits to reduce decoding cost and suppress the accumulation of logical errors~\cite{wang2024rate}. Shorter-depth circuits reduce syndrome-extraction cycles, lower decoding latency, and improve the logical error rate. Thus, compilation quality will remain a key performance lever even as hardware becomes fault-tolerant~\cite{wilson2021empirical}.

\begin{figure}[t]
  \centering
  \includegraphics[width=0.99\columnwidth]{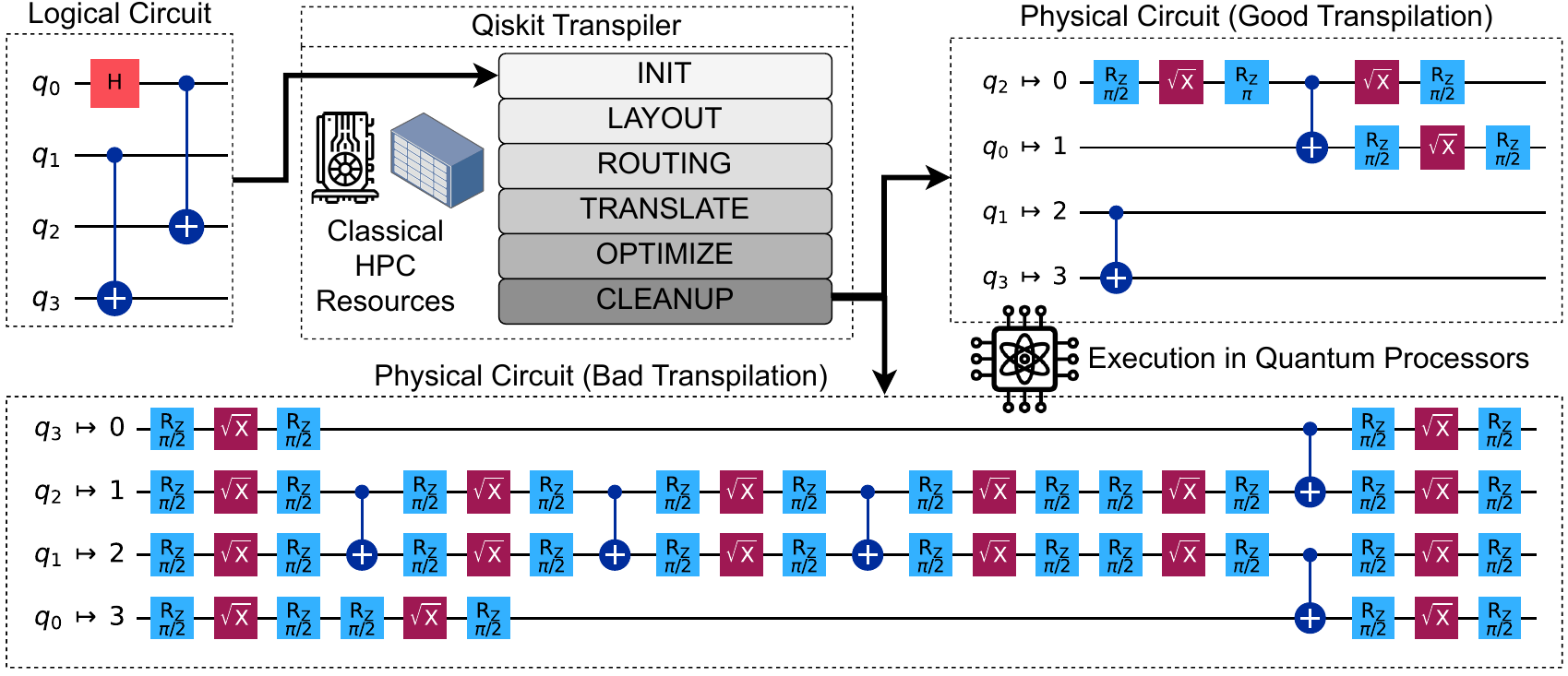}
  \vspace{-4mm}
  \caption{A logical circuit traverses multiple compilation stages, each offering multiple pass options with different algorithmic tradeoffs. The resulting physical circuit quality varies with the selected passes. Both circuits are produced by Qiskit itself -- the compact one with Qiskit (Fidelity Optimized), the inflated one with Qiskit (Time Optimized), and a reversed initial layout (experimental details in Sec.~\ref{sec:implementation}).}
  \vspace{-6mm}
  \label{fig:TuniQ_background}
\end{figure}

At the same time, current quantum hardware is noisy~\cite{kim2023evidence}. Every gate introduces errors, qubit states degrade through decoherence, and error rates vary across qubits within a device. A compiled circuit with more gates accumulates more errors, and greater depth exposes qubits to decoherence for longer. Both effects compound multiplicatively. Fig.~\ref{fig:TuniQ_background} illustrates this. The same four-qubit logical circuit, compiled through different pass sequences, produces physical circuits of vastly different size. The compact circuit (top right) accumulates less noise during execution. The deeper circuit (bottom) contains far more operations, which degrade output fidelity. Compilation quality thus directly determines whether quantum computation produces useful output.

Apart from execution cost, compilation time is equally important for HPC-quantum workloads. For example, HPC routines frequently include variational quantum algorithms~\cite{cerezo2021variational} that recompile circuits with different parameter values across iterations. These algorithms span molecular simulation (VQE~\cite{peruzzo2014variational}), combinatorial optimization (QAOA~\cite{shaydulin2024evidence}), quantum machine learning~\cite{anagolum2024elivagar}, etc. Compilation latency directly limits quantum kernel throughput, and when subroutines are dispatched at scale, the classical compilation step becomes a bottleneck in the classical HPC-quantum pipeline.

\vspace{2mm}

\noindent \textbf{The compiler selection problem.}
Quantum compilation frameworks organize transpilation into multiple sequential stages, each offering several algorithmic options called \emph{passes} (Fig.~\ref{fig:TuniQ_background}). With over 100 passes (and growing), the space of compilation paths is in the range of millions~\cite{li2019tackling, voichick2023qunity}. Transpilers, including the current state-of-the-art from IBM Qiskit~\cite{qiskit2024}, apply a fixed pass sequence uniformly to every circuit. This one-size-fits-all approach ignores that different circuits have different structures and benefit from different passes, that different backends have distinct topologies and error profiles~\cite{ravi2023navigating}, and that noise on a single device drifts over time~\cite{klimov2018fluctuations}. A pass that improves fidelity for one circuit may degrade another. Equally, an ill-chosen sequence can spend significant compilation time on classical HPC resources, while providing no fidelity benefit, or even reducing output quality. The optimal pass selection varies across circuits, backends, and noise conditions simultaneously.

\vspace{2mm}

\noindent \textbf{Research gap.}
Prior work in quantum compilation has developed a rich set of compilers and optimization techniques targeting individual stages of the pipeline~\cite{li2019tackling, murali2019noise, tannu2019not, li2022paulihedral, das2023imitation, gokhale2020optimized, xu2025optimizing, quetschlich2023compiler}.
These improve specific stages but do not address how choices at one stage affect the effectiveness of passes at subsequent stages. Decisions made early in the pipeline propagate and constrain what later stages can achieve, so optimizing stages in isolation produces globally suboptimal circuits.
Recent work further confirms that aggressive pass application often yields negligible fidelity gains~\cite{huo2025revisiting}. Yet no system selects passes adaptively across all stages in a unified framework, and exhaustive search over the full configuration space is intractable. In our experience, simple approaches like greedy per-stage optimization also fail because locally optimal choices at individual stages produce worse end-to-end circuits.

\vspace{2mm}

\noindent \textbf{Our approach: \sol{}.}
We present \sol{}, a \emph{selector} over existing compilation passes that determines which to apply and which to skip at each stage -- adapting to the input circuit, target backend, and current noise profile. As we discuss in Sec.~\ref{sec:design}, the requirements of this problem naturally suit a reinforcement learning (RL) pipeline. \sol{} formulates pass selection as an RL problem where an agent learns from end-to-end compilation outcomes, which pass decisions lead to better final circuits. A dual-encoder architecture captures logical qubit interactions before hardware mapping and incorporates device noise characteristics after mapping, to enable noise-aware pass selection without retraining as calibrations drift. Shaped intermediate rewards provide credit assignment across stages, helping the agent to learn how early decisions affect downstream effectiveness. \sol{} also performs dynamic action masking to enforce stage dependencies, ensuring a valid, executable transpiled circuit. 

\vspace{2mm}

\noindent \textbf{The following are our major contributions:}

\vspace{1mm}

\noindent (1) \sol{} is, to our knowledge, the first noise-conditioned system to formulate transpilation as a cross-stage pass selection problem, to jointly optimize for circuit fidelity and compilation time. 

\vspace{1mm}

\noindent  (2) \sol{} introduces an RL-based framework with novel extensions, including a dual-encoder for stage-aware representation, shaped rewards for cross-stage credit assignment, and dynamic action masking for guaranteed valid compilation.

\vspace{1mm}

\noindent (3) \sol{} is evaluated extensively with diverse quantum workloads and circuits on real IBM Quantum Cloud processors. It shows improvement in output fidelity and compilation time by an average of 20\% and 34\%, respectively, on representative benchmarks over the current state-of-the-art quantum compiler from IBM Qiskit.

\vspace{1mm}

\noindent (4) \sol{} maintains its effectiveness as hardware changes and noise levels improve over time. Trained only on small circuit instances, \sol{} scales to utility-scale circuits with growing advantage over the state-of-the-art, as circuits grow larger.

\sol{}'s framework and implementation is \textit{open-sourced} for community reuse at: \hyperlink{https://zenodo.org/records/19969999}{https://zenodo.org/records/19969999}. We hope that \sol{} helps the quantum-HPC systems community to build on adaptive compilation pass selection frameworks.
\vspace{-2mm}
\section{Background and Motivation of \sol{}}
\label{sec:motivation}

Quantum computers process information using qubits, two-level quantum systems that can exist in superpositions of basis states and exhibit entanglement with no classical analog. Computation proceeds by applying quantum gates (unitary transformations) to evolve qubit states, followed by measurement that collapses superpositions probabilistically. Quantum programs are expressed as circuits: sequences of gates applied to specific qubits. Among physical implementations, superconducting transmon qubits have become dominant due to compatibility with semiconductor fabrication, fast gate operations, and sufficient coherence time~\cite{koch2007charge, arute2019quantum}.


The circuits that programmers write cannot execute directly on hardware. As introduced in Sec.~\ref{sec:intro}, devices differ in native gate sets and coupling topologies, so circuits need to be translated into hardware-native operations and mapped onto physical qubits. Beyond satisfying these constraints, compilation should also minimize gate count and circuit depth to reduce noise exposure and, where possible, transform circuits to be more resilient to errors. This process, called \textit{transpilation}, is performed on classical high-performance resources such as GPUs.


\vspace{2mm}

\noindent\textbf{Transpilation pipeline.} Fig.~\ref{fig:TuniQ_background} illustrates this process using Qiskit, the dominant open-source quantum computing framework~\cite{javadi2024quantum}. Qiskit organizes transpilation into six sequential stages. The \emph{init} stage decomposes multi-qubit gates into one- and two-qubit primitives. The \emph{layout} stage maps logical qubits to physical qubits on the device. The \emph{routing} stage inserts SWAP gates so that all two-qubit interactions occur between adjacent physical qubits. The \emph{translate} stage converts gates into the target device's native basis set. The \emph{optimize} stage cancels redundant gates and consolidates sequences. Finally, the \emph{cleanup} stage removes trivial operations introduced by earlier stages. Each stage offers multiple \emph{transpiler passes}, which are modular transformations that implement different algorithms with distinct tradeoffs. A layout pass might prioritize speed or search exhaustively for higher-quality mappings. A routing pass might greedily insert SWAPs or use lookahead heuristics. The choice of passes at each stage compounds: six stages with several options each, yielding a combinatorial space of transpilation paths~\cite{sivarajah2021t}. As Fig.~\ref{fig:TuniQ_background} shows, the same circuit can produce physical circuits of very different sizes/ quality depending on which path the compiler takes.

\vspace{2mm}

\begin{figure}[t]
  \centering
  \includegraphics[width=0.99\columnwidth]{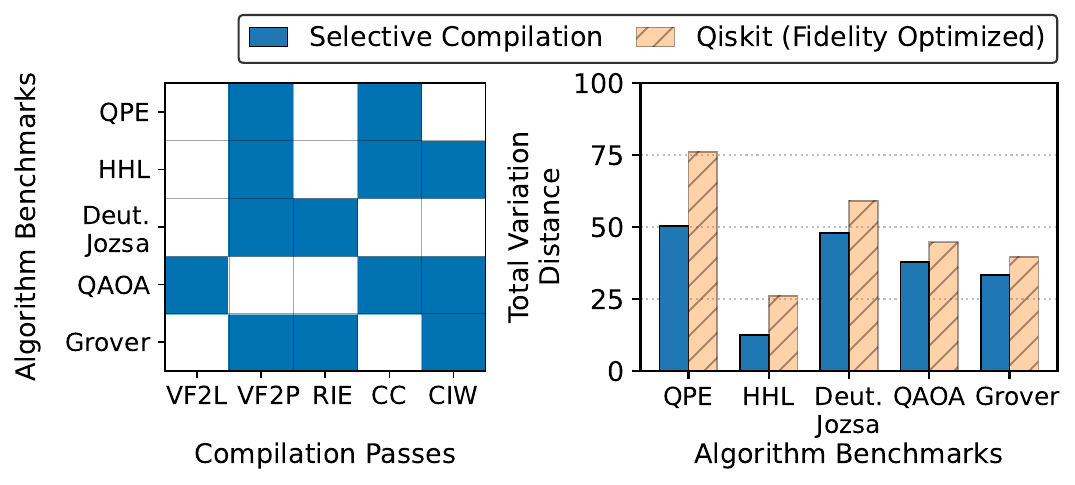}
  \vspace{-4mm}
  \caption{Optimal pass selection (selective compilation) varies by circuits (blue box means pass is enabled, white box means that it is disabled (left). Selective compilation consistently outperforms Qiskit (Fidelity Optimized) (right).}
  \vspace{-6mm}
  \label{fig:motivation_different_circuits}
\end{figure}

\noindent\textbf{Noise and fidelity.} What constitutes a ``good'' transpilation differs fundamentally from classical compilation. Classical compilers optimize primarily for execution speed; quantum compilers optimize for \emph{fidelity} -- the probability that the output matches ideal, noise-free execution. This distinction arises because current quantum hardware operates in the noisy intermediate-scale quantum (NISQ) regime, where every operation introduces errors and quantum states degrade continuously over time~\cite{preskill2018quantum}.

Gate operations fail with probability $\epsilon_g$, where two-qubit gates exhibit error rates of $10^{-3}$ to $10^{-2}$, roughly 10--100$\times$ worse than single-qubit operations~\cite{kim2023evidence}. Qubits also lose information through decoherence: relaxation time $T_1$ governs energy decay while dephasing time $T_2$ governs phase randomization~\cite{krantz2019quantum}. These parameters vary across qubits and edges within a device and drift as calibrations change over hours to days. A circuit with more gates accumulates more gate errors; a circuit with greater depth exposes qubits to decoherence for longer. Both effects compound multiplicatively, so even modest reductions in gate count or depth can substantially improve output quality and the effectiveness of the computation.

The key metric to quantify fidelity is Total Variation Distance (TVD) between the ideal output distribution $p$ and the observed distribution $q$: $\text{TVD}(p,q) = \frac{1}{2}\sum_x |p(x) - q(x)|$, where lower values indicate higher fidelity~\cite{lubinski2023application}. However, measuring TVD requires executing circuits on quantum hardware, which is costly and impractical during compilation when many candidate circuits needs to be evaluated. Hence, we can also use Estimated Success Probability (ESP) as a pre-execution proxy: $\text{ESP} = \prod_{i}(1 - \epsilon_{g_i}) \cdot e^{-d \cdot t_g / T_1} \cdot e^{-d \cdot t_g / T_2}$, where $\epsilon_{g_i}$ is the error rate of gate $i$, $d$ is circuit depth, and $t_g$ is average gate duration~\cite{tannu2019not}. ESP captures the dominant physical error mechanisms like gate failures and decoherence, using calibration data. It is a lightweight proxy that correlates with measured TVD.

\vspace{2mm}

\noindent\textbf{The pass selection challenge.} Not all transpiler passes are required for correct compilation. A minimal pass sequence satisfies hardware constraints -- decomposing gates, mapping qubits, routing interactions, and translating to the native basis, but produces unoptimized circuits. The majority of passes are optional: they reduce gate count, improve qubit mapping on hardware, or restructure circuits to improve fidelity. Qiskit provides over 100 transpiler passes across its six stages and bundles subsets into optimization levels 0 through 3. Level 3~\cite{javadi2024quantum}, which we refer to as \emph{Qiskit (Fidelity Optimized)} (state-of-the-art transpiler), represents the most aggressive configuration. It enables the largest set of optimization passes and is designed to maximize output fidelity. However, even Level 3 applies a fixed set of passes uniformly to all circuits, ignoring circuit-specific structure. With over 100 passes and six stages, the space of possible transpilation paths numbers in the millions, which makes exhaustive exploration intractable~\cite{li2019tackling,voichick2023qunity}.

Fig.~\ref{fig:motivation_different_circuits} illustrates this challenge using five representative passes: \textit{VF2Layout} (VF2L) and \textit{VF2PostLayout} (VF2P) select high-quality qubit mappings via subgraph isomorphism, \textit{RemoveIdentityEquivalent} (RIE) eliminates gates with negligible effect, \textit{CommutativeCancellation} (CC) cancels redundant gates by exploiting commutativity, and \textit{ContractIdleWiresInControlFlow} (CIW) removes unused qubits from control-flow operations. Qiskit (Fidelity Optimized) enables all five passes for every circuit. We compare this against \emph{selective compilation}, which performs brute-force search over all $2^5$ pass combinations to identify the best configuration per circuit for different representative algorithm benchmark circuits (details in Sec.~\ref{sec:implementation}). The heatmap reveals that no single pass combination works best across all benchmarks. The bar chart shows that selective compilation achieves substantially lower TVD than Qiskit (Fidelity Optimized) on every benchmark. This variation arises because pass effectiveness depends on circuit structure. For example, circuits with many commuting gates benefit from CC, while those with complex qubit interactions require careful layout selection via VF2L -- and passes that help one circuit may introduce overhead or interfere with other optimizations in another. These results demonstrate that fixed configurations is not good for fidelity, yet identifying optimal configurations through exhaustive search is impractical at scale.

\begin{observationbox}
\textbf{Observation:} The optimal set of quantum compiler passes varies across circuits, and fixed configurations consistently underperform circuit-specific selection.
\end{observationbox}


\begin{figure}[t]
  \centering
  \includegraphics[width=0.99\columnwidth]{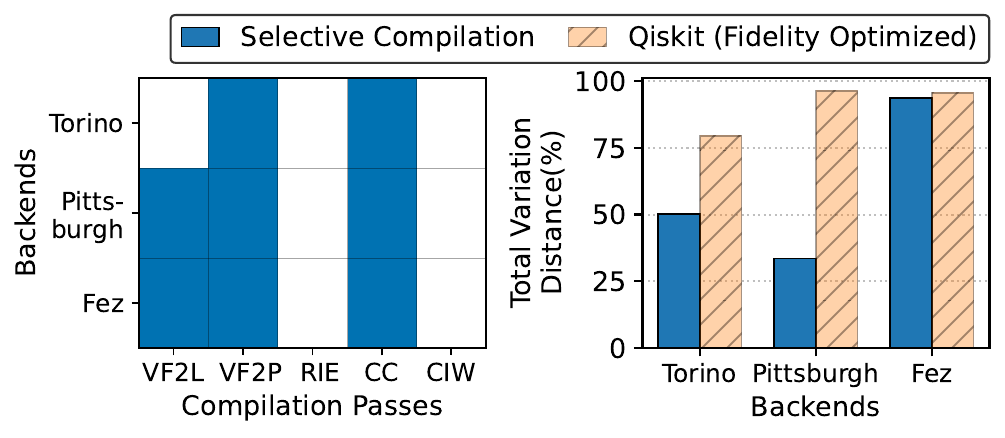}
  \vspace{-4mm}
 \caption{Pass selection varies across backends. Left: optimal passes differ for QPE across three IBM Quantum devices (blue = enabled, white = disabled). Right: selective compilation outperforms Qiskit (Fidelity Optimized) on all backends.}
 \vspace{-6mm}
  \label{fig:motivation_diff_backend}
\end{figure}

\noindent\textbf{Variation across quantum hardware backends.}
The optimal pass configuration varies with the target hardware. Different quantum backends have distinct coupling topologies, native gate sets, and error profiles, so a pass configuration that works well on one device may underperform on another.
Fig.~\ref{fig:motivation_diff_backend} demonstrates this by compiling Quantum Phase Estimation (QPE) on three IBM Quantum Cloud backends: Torino, Pittsburgh, and Fez (experimental details in Sec.~\ref{sec:implementation}). Each backend requires a different subset of passes to achieve optimal fidelity via selective compilation. The bar chart confirms that selective compilation is better on all backends.

\vspace{1mm}

\begin{observationbox}
\textbf{Observation:} The optimal pass configuration depends on the target hardware, including its coupling topology, native gate set, and error profile, which vary across devices.
\end{observationbox}

\vspace{1mm}

\begin{figure}[t]
  \centering
  \includegraphics[width=0.99\columnwidth]{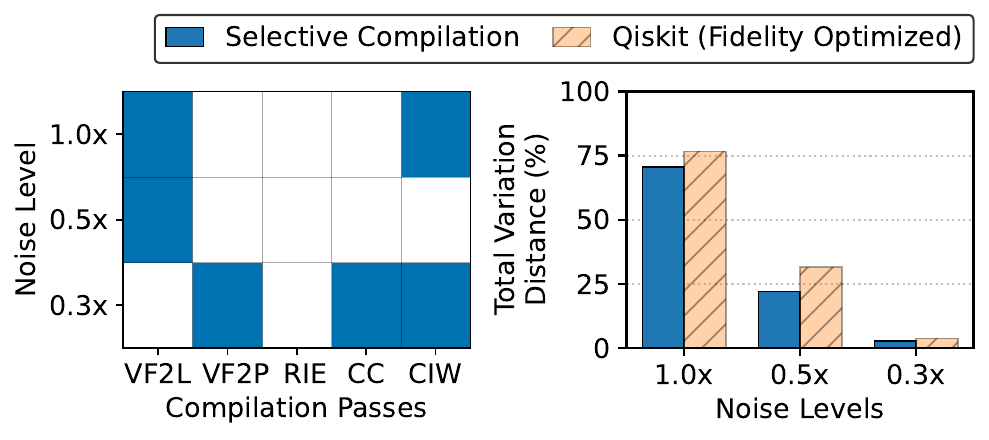}
  \vspace{-4mm}
  \caption{Selective compilation achieves lower TVD for Deutsch-Jozsa as device noise scales from baseline (1.0$\times$) to reduced (0.3$\times$). (blue = enabled, white = disabled))}
  \vspace{-6mm}
  \label{fig:motivation_noise_change}
\end{figure}

\noindent\textbf{Variation across noise conditions.}
Beyond backend differences, noise levels on a single device drift over time as calibrations change and environmental conditions fluctuate. A compiler strategy that assumes static noise will produce suboptimal results as the hardware evolves.
Fig.~\ref{fig:motivation_noise_change} shows this by compiling another representative benchmark, Deutsch-Jozsa, under varying noise conditions. We scale IBM Torino's calibration data (readout and gate errors) by factors of 1.0$\times$, 0.5$\times$, and 0.3$\times$ to simulate different noise regimes. As noise decreases, the optimal pass set changes. This shift occurs because the relative importance of different optimizations depends on the noise regime: under high noise, layout quality and reducing idle time dominate since every additional gate or delay accumulates significant error; as noise decreases, post-layout refinement becomes worthwhile because the marginal gains from better qubit placement outweigh the overhead of additional optimization.

\begin{observationbox}
\textbf{Observation:} The optimal pass configuration depends on current noise characteristics, which drift over time on any device.
\end{observationbox}

\noindent Based on these observations, next, we discuss the design of \sol{}.


\section{System Design of \methodname}
\label{sec:design}

The observations in Sec.~\ref{sec:motivation} establish that optimal pass selection varies with circuit structure and hardware conditions. \methodname{} addresses this using reinforcement learning to select transpiler passes that improve execution fidelity, while reducing compilation time.

\subsection{\sol{}'s Objectives}

\noindent\textbf{Fidelity.} The primary goal is to maximize the probability that a compiled circuit produces correct output on noisy hardware. We quantify fidelity using TVD during evaluation, but TVD requires executing circuits on quantum hardware and comparing against an ideal simulation. This is infeasible during compilation, where many candidate pass configurations need to be evaluated. We therefore use ESP as a training proxy. ESP captures the dominant physical error mechanisms: gate failures, readout errors, and decoherence, using only calibration data, and in our experience, correlates well with measured TVD (Sec.~\ref{sec:motivation} defines TVD and ESP).

\vspace{1mm}

\noindent\textbf{Compilation time.} A key feature of quantum workloads is that they frequently recompile the same circuit structure with different parameters. For example, variational algorithms~\cite{cerezo2021variational} iterate thousands of times, each iteration requiring a fresh compilation as gate angles change. Near-term applications in chemistry and optimization~\cite{farhi2016quantum,peruzzo2014variational} spend substantial time in this compile-execute loop. A compiler that improves fidelity but increases compilation time may yield no net benefit if the application is compilation-bound. \methodname therefore constrains compilation time to remain competitive with Qiskit (Fidelity Optimized), to make sure that fidelity gains do not come at the cost of throughput of circuit execution.

\subsection{Pass Selection Decisions}

Achieving these objectives requires identifying which compilation decisions to optimize. We focus on \emph{pass selection}: choosing which transpiler passes to apply at each stage. As shown in Sec.~\ref{sec:motivation}, not all passes benefit every circuit -- applying unnecessary passes can incur overhead or even degrade fidelity, while omitting beneficial passes results in suboptimal output quality. We organize decisions across five stages of the Qiskit transpilation pipeline. The \textit{init} stage decomposes multi-qubit gates into primitives and applies optional early optimizations on the logical circuit. The \textit{layout} stage maps logical qubits to physical qubits. This decision determines which physical error rates affect each operation and how many SWAP gates routing will be required, with algorithms like VF2Layout~\cite{cordella2004sub} exposing parameters such as call limits and trial counts. The \textit{routing} stage inserts SWAP gates to satisfy connectivity constraints~\cite{li2019tackling}, directly affecting circuit depth, gate count, and which physical edges are utilized. The \textit{translation} stage converts gates into the target hardware's native gate set using standard decomposition rules, ensuring the circuit is physically executable. The \textit{optimization} stage selects a subset of passes and applies them iteratively until convergence, exposing different opportunities for gate cancellation and resynthesis. The \textit{cleanup} stage applies lightweight passes to remove residual redundancies introduced by earlier transformations.

Passes across stages interact: layout choices constrain routing options, routing outcomes determine which optimizations are effective, and early decisions propagate through subsequent stages. Independent tuning of each stage fails because locally optimal choices may be globally suboptimal. For instance, on QPE compiled for IBM Torino, selecting the best layout pass (VF2L) and best optimization pass (CC) independently yields a TVD of 62\%, while jointly selecting VF2L without CC achieves 51\%. The individually optimal CC pass undoes gate arrangements that VF2L's qubit placement leverages. \textit{Pass selection should therefore not be performed in isolation; joint optimization across stages is essential.}

\subsection{Design Requirements}
\label{sec:des_req}
Jointly optimizing pass selection under diverse circuit and hardware characteristics imposes several requirements. \req{1} The policy should generalize across circuits with varying sizes, gate compositions, and connectivity patterns; training on one distribution (e.g., random circuits) should transfer to another (e.g., algorithmic benchmarks). \req{2} The policy should adapt to hardware characteristics, including coupling topology and current noise conditions, since calibrations drift over hours to days~\cite{klimov2018fluctuations}. \req{3} Decisions should be stage-aware: early passes constrain later options -- \textit{e.g.}, choosing a layout pass requires anticipating its impact on routing and optimization, not just evaluating immediate effect. \req{4} Inference overhead should be negligible compared to transpilation time.

Beyond performance, the selected pass sequence should produce circuits that execute correctly. Not all combinations yield valid output: skipping required routing leaves connectivity unsatisfied, while certain orderings produce gates outside the target basis. The policy should propose sequences that satisfy hardware constraints and preserve functional equivalence. We found that reinforcement learning satisfies these design requirements. Before providing design details, we discuss the reasoning behind this choice.

\subsection{Why Reinforcement Learning?}
\label{sec:whyrl}

\begin{table}[t]
\centering
\caption{Compiled circuit metrics for supervised DNN versus vanilla RL on QPE benchmark. Lower is better; fewer two-qubit gates and shallower depth reduce noise accumulation.}
\label{tab:sl_vs_rl}
\vspace{-4mm}
\begin{tabular}{lcc}
\toprule
Method & 2Q Gate Count & Depth (post compilation) \\
\midrule
Supervised (DNN) & 1244 & 1786 \\
Vanilla RL (DQN) & 327 & 742 \\
\bottomrule
\vspace{-6mm}
\end{tabular}
\end{table}

\begin{figure}[t]
  \centering
  \includegraphics[width=0.99\columnwidth]{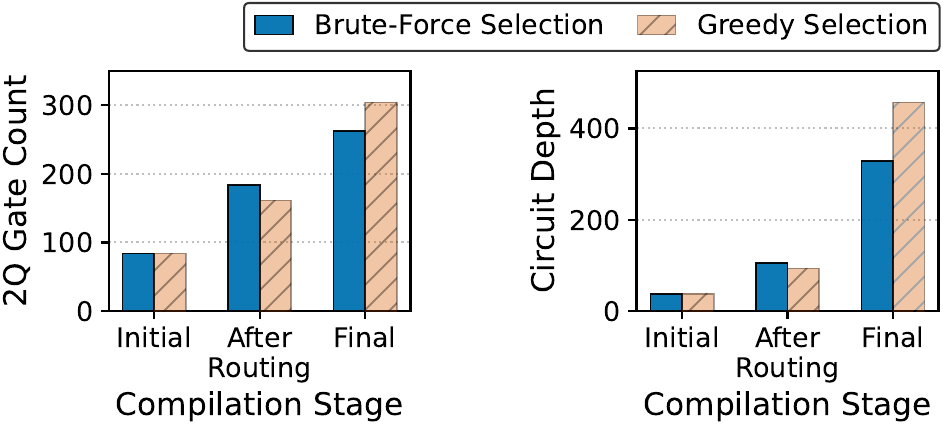}
  \vspace{-3mm}
  \caption{Greedy pass selection achieves better intermediate metrics but worse final circuits. Effective pass selection requires optimizing for end-to-end quality.}
  \vspace{-4mm}
  \label{fig:design_greedy}
\end{figure}


Fixed strategies like Qiskit's preset levels apply uniform sequences regardless of circuit structure or noise, violating \req{1} and \req{2}. Exhaustive search is infeasible: with $|P_i|$ passes at stage $i$, the space grows as $\prod_i |P_i|$, expanding more when optimization passes iterate.

\begin{figure*}[!tb]
  \centering
  \includegraphics[width=\textwidth]{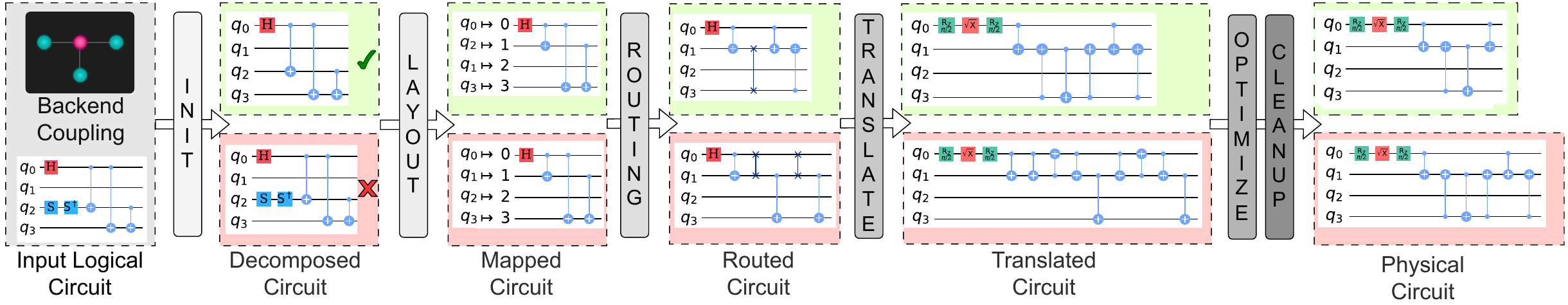}
  \vspace{-8mm}
  \caption{The transpilation pipeline: A logical circuit progresses through six stages, with pass choices at each stage leading to different outcomes. Green boxes indicate high-quality intermediate circuits; red boxes indicate suboptimal choices that accumulate noise. The goal of \sol{} is to select passes to navigate toward the green path, producing a compact physical circuit.}
  \label{fig:design_overview}
  \vspace{-4mm}
\end{figure*}

A natural approach is supervised learning: extract successful pass sequences from selective compilation, pair them with circuit features, and train a classifier. Table~\ref{tab:sl_vs_rl} compares a supervised neural network against a vanilla Deep Q-Network (DQN), a reinforcement learning algorithm that learns to select actions by estimating their long-term value. Both models use identical architectures trained on random circuits (5--10 qubits) and evaluated on a 13-qubit representative QPE benchmark. Two-qubit (2Q) gate count and circuit depth serve as fidelity proxies since two-qubit gates exhibit 10--100$\times$ higher error rates than single-qubit operations, and deeper circuits expose qubits to decoherence longer. The supervised model produces circuits with 3.8$\times$ more two-qubit gates and 2.4$\times$ greater depth. This gap arises because supervised learning requires fixed ground-truth labels, but optimal pass sequences shift as circuits and noise conditions change. The labels that worked for training circuits may not transfer to evaluation circuits. Additionally, supervised learning treats each stage as an independent classification problem, predicting passes without considering how earlier selections constrain later effectiveness. RL sidesteps both issues: it learns from final compilation outcomes instead of fixed labels, and its reward signal naturally captures cross-stage interactions.

A simpler approach selects passes that greedily optimize a proxy metric at each stage. Fig.~\ref{fig:design_greedy} compares greedy selection (choose passes that minimize ESP at each stage) against brute-force selection (choose passes that minimize final circuit metrics) across seven init-stage optimization passes. Both start from the same initial QPE logical circuit. After routing, greedy selection achieves fewer two-qubit gates and lower depth. Yet, produces worse final circuits. This occurs because passes interact across stages: an init-stage configuration that yields fewer gates after routing may leave the circuit in a form that downstream optimization passes cannot improve further, while a locally suboptimal choice may enable more effective transformations later. What matters is final circuit quality, not intermediate metrics. This sequential structure with delayed consequences aligns naturally with reinforcement learning. An RL agent receives reward signals from final compilation outcomes, not intermediate metrics, learning which early-stage decisions lead to better end results. The agent reasons about downstream effects at each decision point and adapts to circuit features and noise conditions. We now describe how \methodname instantiates this formulation.

\subsection{RL Formulation for Circuit Transpilation}

The above analysis (Sec.~\ref{sec:whyrl}) establishes that effective pass selection requires reasoning about delayed consequences: decisions at early stages shape the effectiveness of later transformations, and optimizing each stage independently produces suboptimal results. We formulate compilation as a Markov Decision Process $\mathcal{M} = (\mathcal{S}, \mathcal{A}, T, R, \gamma)$ where the \emph{agent} is the pass selector, the \emph{environment} is the Qiskit transpiler, and each \emph{episode} corresponds to compiling one circuit. The agent observes the current circuit state, selects a transpiler pass, and the environment applies that pass to produce a new circuit state. This interaction continues until compilation completes, at which point the agent receives a reward based on the quality of the final physical circuit.

Fig.~\ref{fig:design_overview} illustrates the compilation trajectory. A \emph{logical circuit} enters the pipeline and progresses through stages: the init stage produces a \emph{decomposed circuit} containing only primitive one- and two-qubit gates; the layout stage produces a \emph{mapped circuit} with logical qubits assigned to physical qubits; the routing stage produces a \emph{routed circuit} with SWAP gates inserted to satisfy connectivity constraints; and the optimization and cleanup stages produce the \emph{final physical circuit} ready for hardware execution. The circuit representation changes fundamentally after layout on hardware. Before layout, the circuit operates on logical qubits with no hardware binding; after routing, it operates on physical qubits with associated error rates and coherence times. \textit{Our formulation captures this distinction through a dual-encoder architecture, and the sequential decision structure with shaped rewards enables reasoning about how early choices affect downstream outcomes.}

\subsubsection{State Space}

The state $s \in \mathcal{S}$ encodes everything the agent needs to make informed pass selections: what the circuit looks like, what stage of compilation we are in, and what the target hardware looks like. We decompose the state into three components.

\vspace{1mm}

\noindent\textbf{Stage indicator.} A one-hot vector $\mathbf{s} \in \{0,1\}^6$ encodes the current compilation stage (total 6 stages). This enables to learn stage-specific strategies, addressing \req{3} design requirement (Sec.~\ref{sec:des_req}).

\vspace{1mm}

\begin{figure}[!]
  \centering
  \includegraphics[width=\columnwidth]{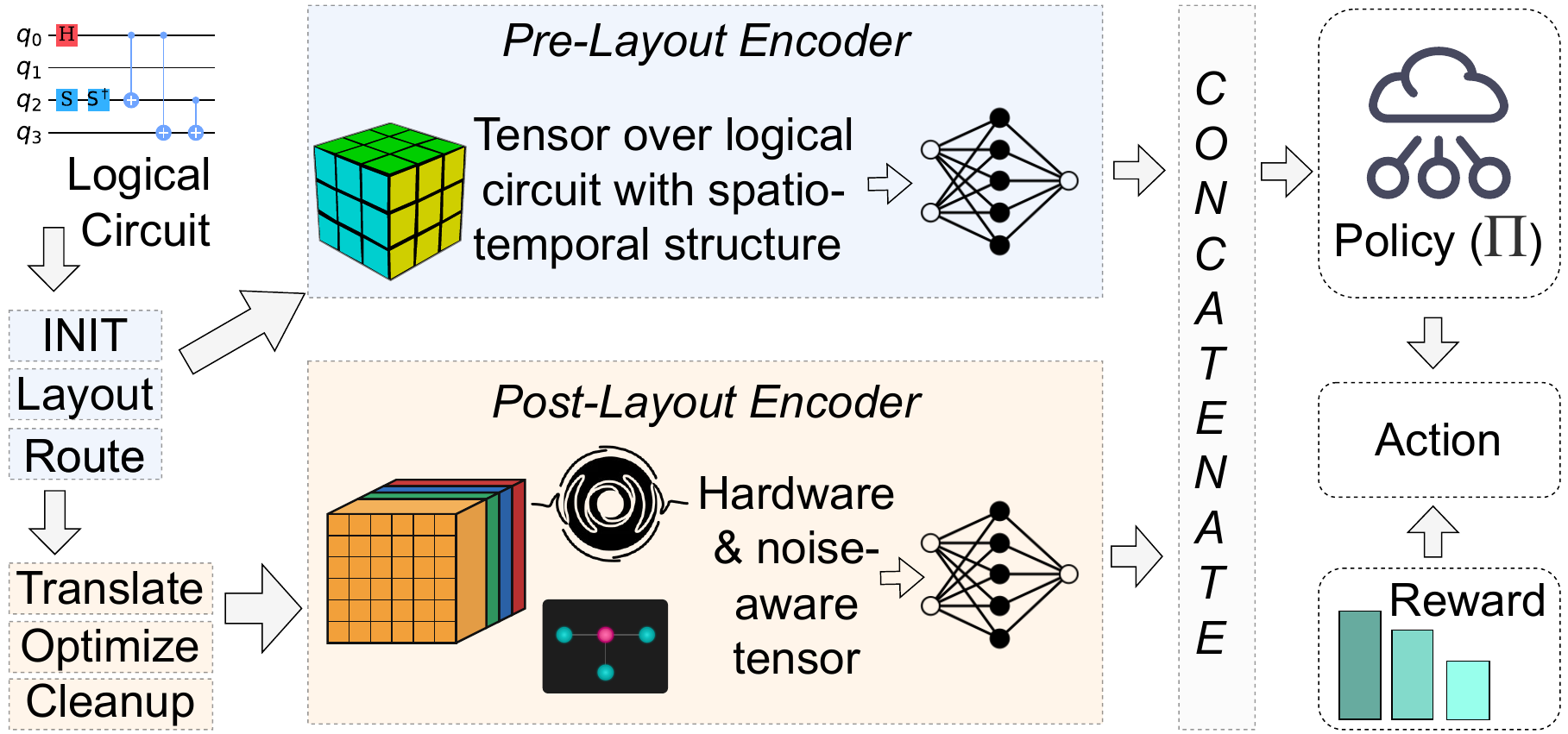}
  \vspace{-6mm}
  \caption{\sol{} employs a dual-encoder architecture: pre-layout encodes spatio-temporal structure over logical qubits; post-routing encodes coupling topology and noise over physical qubits. Concatenated with stage indicators and global features, these enable hardware-aware pass selection.}
  \label{fig:dual-encoder}
  \vspace{-6mm}
\end{figure}

\noindent\textbf{Circuit features.} Before the layout stage, qubits are logical entities with no associated noise characteristics. After routing, every gate operates on specific physical qubits with calibrated error rates and coherence times. A single encoder cannot capture both representations effectively. We design a \emph{dual-encoder architecture} that switches representation based on the compilation stage (Fig.~\ref{fig:dual-encoder}).

For pre-layout stages (init, layout, routing), we encode the circuit as tensors over logical qubits. Two-qubit gates are represented as $\mathbf{F}^{2q} \in \mathbb{R}^{Q \times Q \times G \times T}$, where $Q$ is the maximum qubit count, $G$ is the number of two-qubit gate types, and $T$ is maximum circuit depth. Entry $F^{2q}_{i,j,g,t} = 1$ if a two-qubit gate of type $g$ connects qubits $i$ and $j$ at time step $t$, and 0 otherwise. This tensor captures the spatio-temporal structure of qubit interactions, important for layout. Single-qubit gates use $\mathbf{F}^{1q} \in \mathbb{R}^{Q \times G' \times T}$ with analogous semantics.

For post-routing stages (optimization, cleanup), the circuit is bound to physical hardware. We augment the tensor with noise characteristics. Two-qubit gates become $\mathbf{F}^{2Q} \in \mathbb{R}^{E \times G \times T \times 4}$, where $E$ is the number of coupling edges and four channels encode: gate count, two-qubit error rate $\epsilon_{2Q}(e)$ for edge $e$, average $T_1$ coherence, and average $T_2$ coherence of the connected qubits. Single-qubit gates become $\mathbf{F}^{1Q} \in \mathbb{R}^{P \times G \times T \times 5}$, where $P$ is the physical qubit count and five channels encode gate count, single-qubit error rates, $T_1$, and $T_2$. This noise-augmented representation enables the agent to learn hardware-aware optimization, addressing design requirement \req{2} by incorporating current calibration data into observations.

Each encoder is a multi-layer perceptron $\phi: \mathbb{R}^{d_{in}} \rightarrow \mathbb{R}^{256}$ with layers $2048 \rightarrow 1024 \rightarrow 512 \rightarrow 256$, ReLU activations, and 10\% dropout. We chose MLPs over graph neural networks because the tensor encoding already captures the spatial and temporal structure of gate interactions, and MLPs evaluate in constant time regardless of circuit size, satisfying \req{4}. The gradual dimension reduction preserves gate-level detail while compressing to a manageable embedding; aggressive compression loses information, while larger embeddings slow inference without improving quality. Experiments with alternative architectures (shared encoder, GNNs, smaller MLPs) produced circuits with overall higher TVD. The pre-layout encoder $\phi_{pre}$ processes logical structure; the post-routing encoder $\phi_{post}$ processes physical characteristics. Separate encoders allow specialization: $\phi_{pre}$ learns qubit interaction patterns relevant to layout, while $\phi_{post}$ learns noise-aware features relevant to optimization.

\vspace{1mm}

\noindent\textbf{Global features.} Circuit-level statistics complement local gate-level encodings: total gate count, circuit depth, parallelism metrics, and topology descriptors of both the circuit interaction graph and backend coupling graph. We include compatibility ratios between circuit and backend topologies to indicate mapping difficulty. These features provide context that local tensor encodings may miss.  Next, we discuss the action space of our RL-based formulation

\subsubsection{Action Space and Masking}
\label{sec:mask}

The action $a \in \mathcal{A}$ corresponds to selecting one transpiler pass from the stage-appropriate pass set, or a \emph{skip} action that advances to the next stage. The action space is discrete and varies by stage: layout passes are available only during layout, routing passes only during routing, and so forth. After selecting a pass, the environment applies it to the current circuit, producing a new state $s'$. The agent then decides whether to apply another pass in the same stage or skip to the next. This continues until all stages are complete. Not all pass sequences produce valid circuits. Skipping required decomposition leaves multi-qubit gates that hardware cannot execute. Skipping routing leaves connectivity constraints unsatisfied. Certain orderings produce gates outside the target basis. We enforce validity through \emph{dynamic action masking} that adapts to current circuit state and compilation progress. 

During initialization, the mask requires decomposing all gates with more than 2 qubits before allowing stage advancement. During layout, it restricts actions to layout algorithms; if an algorithm fails to find a valid mapping within its search limit, the mask removes it and forces selection of an alternative. During routing, the mask ensures connectivity constraints are satisfied before proceeding. After routing, optimization passes become available only after basis translation completes; the mask tracks which gates remain outside the target basis and forces translation when necessary. Some passes require specific follow-up passes; the mask enforces these dependencies. \textit{This mechanism guarantees that every completed episode produces a valid, hardware-executable circuit. The agent learns \emph{which} valid passes to select; validity is guaranteed by construction.}

\subsubsection{Transition Dynamics}

When the agent selects an action, the environment executes the corresponding pass on the current circuit. The transition function $T(s' | s, a)$ is deterministic: given circuit state $s$ and action $a$, applying pass $a$ produces a unique next state $s'$. The Qiskit transpiler serves as the environment, executing passes and returning updated circuit states. For passes with stochastic behavior (e.g., SabreSwap uses random tie-breaking), we fix seeds to make transitions reproducible during training. After each transition, the agent receives a reward signal that quantifies how the action affected compilation quality, as we discuss next.

\subsubsection{Reward Structure}

The reward function $R(s, a, s')$ quantifies how much each action improves compilation quality. Assigning a reward only at episode completion provides a weak credit assignment signal: compilation spans multiple stages with dozens of decisions, and a sparse reward cannot distinguish which choices contributed to the outcome. We use \textit{shaped rewards} that provide intermediate feedback while preserving the importance of final quality, directly supporting \req{3}. We define a \emph{transpilation quality} metric TQ that serves as an ESP proxy throughout compilation, \textcolor{black}{where ESP cannot yet be computed directly}:
\begin{equation}
\label{eq:TQ}
\text{TQ} = \left(\prod_{g \in \mathcal{G}_{1Q}} s_{1Q}(g)\right) \left(\prod_{g \in \mathcal{G}_{2Q}} s_{2Q}(g)\right) \cdot \exp\left(-\frac{d \cdot t_g}{\bar{T}_1} - \frac{d \cdot t_g}{\bar{T}_2}\right)
\end{equation}
where $\mathcal{G}_{1Q}$ and $\mathcal{G}_{2Q}$ are sets of one- and two-qubit gates, $s_{1Q}$ and $s_{2Q}$ are gate success probabilities, $d$ is circuit depth, $t_g$ is average gate duration, and $\bar{T}_1$, $\bar{T}_2$ are coherence times over active qubits.

Computing TQ requires gate success probabilities, which differ before and after routing. During layout, two-qubit gates may connect non-adjacent physical qubits. We estimate routing overhead using shortest-path distance on the coupling graph: for a gate mapped to qubits at distance $d_{path} > 1$, routing requires approximately $d_{path} - 1$ SWAP operations, each decomposing into three two-qubit gates. The effective error becomes $\epsilon_{2Q}^{eff} = (d_{path} - 1) \cdot 3 \cdot \bar{\epsilon}_{2Q} + \bar{\epsilon}_{2Q}$, and predicted depth increases accordingly. This \emph{layout quality} (LQ) metric estimates final circuit quality from layout decisions alone, providing a reward signal before routing occurs. After routing, all two-qubit gates act on adjacent physical qubits, \textcolor{black}{but the circuit is not yet in the target basis, so per gate calibrated error rates are not yet directly applicable.} \emph{Routing quality} (RQ) \textcolor{black}{instead uses per-edge average error rates with decomposition-aware scaling to account for gates that will expand into multiple basis gates after translation.} During optimization and cleanup, the circuit is fully translated \textcolor{black}{and bound to physical qubits, so per-gate error rates are now directly available from calibration data. At this point, ESP can be computed exactly and replaces TQ as the reward signal — which is why the episode-end reward in Eq.~\ref{eq:final_reward} uses ESP rather than TQ. In summary: TQ (via LQ at layout, RQ after routing) provides shaped intermediate rewards when exact gate-level error attribution is unavailable; ESP takes over once translation completes.} The final reward combines ESP improvement with auxiliary metrics:
\begin{equation}
\label{eq:final_reward}
R_{final} = W \cdot \text{clip}\left(\log \frac{\text{ESP}_{rl}}{\text{ESP}_{L3}}\right) + \phi\left(w_1 \cdot r_{gates} + w_2 \cdot r_{depth}\right)
\end{equation}
where $r_{gates} = (gates_{L3} - gates_{rl}) / \max(gates_{L3}, 1)$ and $r_{depth}$ measures relative improvement over Qiskit Level 3 transpilation (Qiskit (Fidelity Optimized)), the most aggressive optimization level and our primary baseline. The logarithmic ratio rewards multiplicative ESP improvements. Auxiliary terms encourage gate count and depth reduction even when ESP gains are marginal.

\begin{algorithm}[t]
\caption{Transpilation with \methodname{}}
\label{alg:compilation}
{\footnotesize
\begin{algorithmic}[1]
\STATE \textbf{Input:} Logical circuit $C$, backend $B$, policy $\pi$, mode $\in$ \{Train, Infer\}
\STATE \textbf{Output:} Physical circuit executable on $B$
\STATE Preprocess $C$: unroll custom gates, remove identities
\STATE $stage \gets 0$ \hfill $\triangleright$ Init $\to$ Layout $\to$ Route $\to$ Translate $\to$ Optimize $\to$ Cleanup
\WHILE{$stage < 6$}
    \STATE Encode $C$ with $\phi_{pre}$ (if $stage \leq 2$) or $\phi_{post}$ using calibration from $B$
    \STATE Construct state $s$: encoding $\oplus$ stage indicator $\mathbf{s}$ $\oplus$ global features $\mathbf{g}$
    \STATE Compute action mask $\mathbf{m}$ \hfill $\triangleright$ Enforce stage constraints
    \STATE $a \gets \begin{cases} \text{sample from } \pi(a \mid s, \mathbf{m}) & \text{if train} \\ \arg\max_a \pi(a \mid s, \mathbf{m}) & \text{if infer} \end{cases}$
    \STATE \textbf{if} $a$ is \texttt{skip}: $stage \gets stage + 1$
    \STATE \textbf{else}:
    \STATE \hspace{1.5em} Apply pass $a$ to $C$ \hfill $\triangleright$ Qiskit executes pass
    \STATE \hspace{1.5em} Compute shaped reward: LQ, RQ, or ESP \hfill $\triangleright$ Train only; stage-dependent
\ENDWHILE
\STATE Compute $R_{final}$ (Eq.~\ref{eq:final_reward}) using Qiskit Level 3 reference \hfill $\triangleright$ Train only
\STATE Update $\pi$ via MaskablePPO \hfill $\triangleright$ Train only
\RETURN $C$
\end{algorithmic}}
\end{algorithm}

\subsubsection{Putting It All Together: Training and Inference}

Alg.~\ref{alg:compilation} summarizes \sol{}'s compilation procedure. Training and inference share the same core loop (state, action masking, pass application via Qiskit) but differ in action selection and reward computation.

\vspace{1mm}

\noindent\textbf{Training.} Given a logical circuit and target backend, the agent preprocesses the circuit to unroll custom definitions. At each stage, it observes state $s$ (circuit features from the stage-appropriate encoder, backend properties, stage indicator), computes action mask $\mathbf{m}$, and samples $a \sim \pi(a|s, \mathbf{m})$. A skip action advances to the next stage; otherwise, the environment applies pass $a$ and returns new state $s'$. Intermediate TQ (Eq.~\ref{eq:TQ}) provides shaped rewards at each transition: LQ during layout, RQ after routing, and ESP during optimization and cleanup. At episode completion, $R_{final}$ (Eq.~\ref{eq:final_reward}) computes the log-ratio of the agent's ESP against a Qiskit Level~3 reference compilation on the same circuit, combined with auxiliary gate count and depth terms. Policy parameters are then updated via MaskablePPO.

\vspace{1mm}

\begin{figure}[!]
  \centering
  \includegraphics[width=\columnwidth]{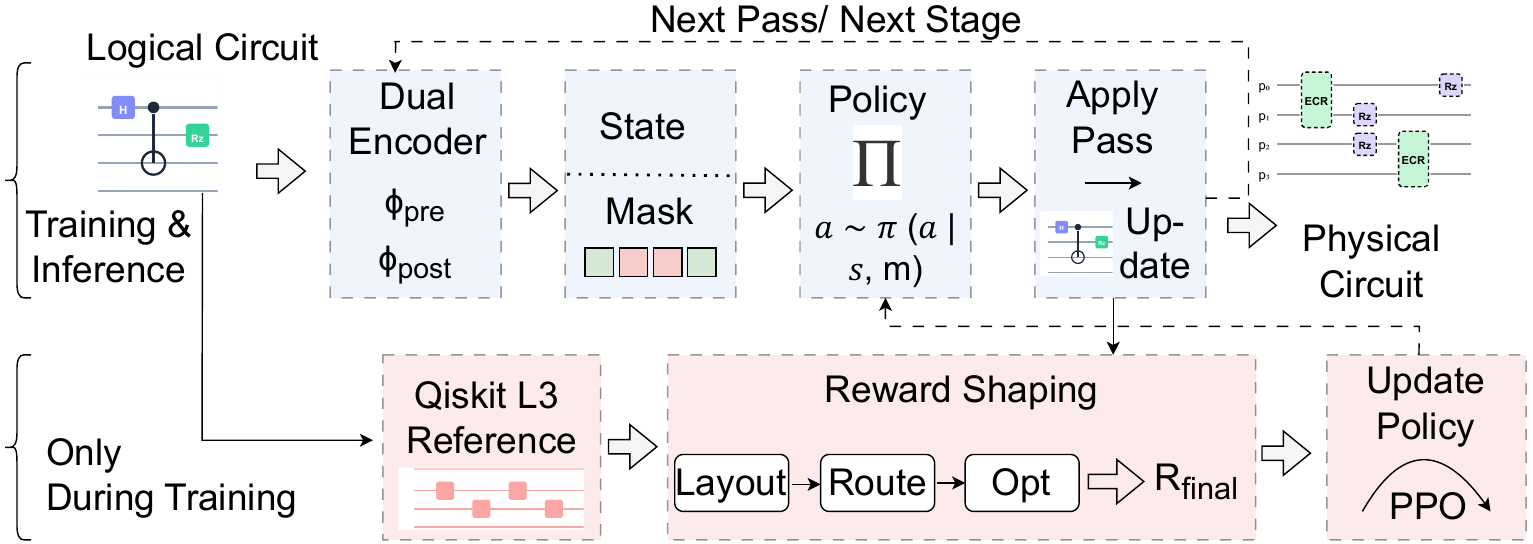}
  \vspace{-6mm}
  \caption{Overview of \sol{}'s compilation pipeline. The shared path (top) is used during both training and inference. Training-only components (bottom) are absent at inference.}
  \label{fig:train_and_infer}
  \vspace{-6mm}
\end{figure}

\noindent\textbf{Inference.} At deployment, \sol{} uses a frozen policy with no reward computation, no policy updates, and no Qiskit Level~3 reference compilation. \textit{ The Qiskit Level~3 reference in Eq.~\ref{eq:final_reward} is required \emph{only} during training; it is entirely absent at inference time.} The loop structure is otherwise identical: \sol{} constructs state $s$ via the dual encoder ($\phi_{pre}$ or $\phi_{post}$ depending on stage), computes mask $\mathbf{m}$, and selects $a = \arg\max_a \pi(a|s, \mathbf{m})$ from the frozen policy. Three components are eliminated relative to training: (1)~stochastic exploration is replaced by greedy action selection; (2)~all reward signals (TQ, LQ, RQ, ESP) and their associated shortest-path, noise-lookup, and ESP computations are removed; and (3)~the per-episode Qiskit Level~3 compilation is not executed. The entire per-step overhead reduces to one MLP forward pass and a masked $\arg\max$, which adds less than 1\% to total transpilation time in our measurements. The compilation pipeline of \sol{} is visually depicted in Fig.~\ref{fig:train_and_infer}.

\subsection{Improving \sol{}'s Robustness}

\textit{Real-world compilation encounters corner cases that can destabilize learning or produce invalid output.} We address these through preprocessing, runtime safeguards, and training diversity.

\vspace{1mm}

\noindent\textbf{Pass failures.} Certain passes fail under specific conditions: \texttt{VF2Layout} may exhaust its call limit without finding a valid mapping; routing algorithms may enter long-running loops on pathological circuits; optimization passes may trigger gate explosion or oscillatory behavior. \sol{} wraps pass execution with configurable timeouts and track failed passes per episode, masking them from future selection. Fallback mechanisms ensure compilation always completes: if all layout algorithms fail, the agent falls back to a trivial layout; if routing times out, a greedy baseline executes instead.

\vspace{2mm}

\noindent\textbf{Stage enforcement.} Dynamic action masking prevents invalid stage transitions. The agent cannot attempt routing before layout completes or optimization before basis translation. At episode end, a basis check verifies all gates are in the target set; if violations exist, forced retranslation corrects them before returning the circuit.

\vspace{2mm}

\noindent\textbf{Reward stability.} ESP values can span orders of magnitude, and pathological circuits may produce near-zero ESP that destabilizes learning. We bound rewards using log-ratio relative to Qiskit Level 3 transpilation (Qiskit (Fidelity Optimized)), to ensure controlled magnitudes regardless of absolute ESP. Auxiliary terms based on gate count and depth provide a meaningful signal even when ESP differences are negligible. \sol{} also penalizes repeated no-op selections -- passes that produce no net gate-count or depth change and terminate optimization stages on convergence. When opposing passes appear (one increases gate count while another decreases it), the agent observes the cumulative net effect through state encoding and shaped rewards; if iterations produce no net improvement, the convergence criterion fires and the loop exits.

\vspace{2mm}

\noindent\textbf{Backend diversity.} Calibration drift and hardware variability can cause policies to overfit to specific noise profiles. During training, we perturb backend instances by scaling gate errors and coherence times ($T_1$, $T_2$) across a range, and randomly disable coupling edges to simulate partial connectivity failures. This exposure ensures the policy generalizes across backend conditions encountered in deployment. Next, we discuss \sol{}'s experimental details. 

\vspace{2mm}

\noindent\textbf{Generalization as hardware and compilers evolve.} The design of \sol{} is agnostic to specific platforms. The state representation encodes circuit structure and hardware characteristics (error rates, coherence times, coupling graphs) fundamental to any quantum technology, whether superconducting qubits, trapped ions, neutral atoms, or other modalities. The post-routing encoder takes these as input features; adapting to new hardware requires only calibration data. Action masking is defined by pass dependencies, so extending to new passes requires only updating mask rules. This ensures the design remains applicable as hardware and compilers evolve and diversify. \sol{} operates at the physical-circuit level (pre-QEC); error-corrected circuits enter \sol{} as compilation targets like any other circuit, with QEC-specific logical compilation composing cleanly as a separate upstream layer.

\section{Implementation and Methodological Details}
\label{sec:implementation}

\methodname{} implements the RL formulation as a Gymnasium environment~\cite{towers2024gymnasium} wrapping Qiskit's transpiler infrastructure. \textit{This enables integration with any policy gradient algorithm supporting action masking, while providing a simple interface for end users.} Here, we describe the system interface, architecture, training configuration, and other experimental details.

\vspace{2mm}

\noindent\textbf{Interface and environment.} \methodname{} exposes an API mirroring Qiskit's transpilation interface for drop-in replacement. Beneath this, the Gymnasium environment comprises four components: \textit{CircuitState} maintains the circuit and applies passes; \textit{FeatureBuilder} constructs observation tensors; \textit{ActionMaskManager} enforces stage constraints; and \textit{RewardCalculator} computes shaped rewards.
\begin{lstlisting}[language=Python, basicstyle=\footnotesize\ttfamily, frame=single, numbers=left, xleftmargin=2em]
# User-facing API (drop-in for qiskit.transpile)
compiler = TuniQCompiler(policy_path="policy.zip", backend=backend)
physical_circuit = compiler.compile(logical_circuit)

# Environment interface for policy training
class TuniQEnv(gymnasium.Env):
    def step(self, action):
        if action == SKIP_ACTION:
            self.stage += 1; self._update_mask()
        else:
            if not self.circuit_state.apply_pass(action):
                self.mask_manager.mark_failed(action)
        return self.feature_builder.extract(), \
               self.reward_calc.compute(), self.stage >= NUM_STAGES
    def action_masks(self):
        return self.mask_manager.get_valid_actions(self.stage)
\end{lstlisting}
\textit{This separation enables independent testing and integration with any policy gradient algorithm supporting action masking.} Batch compilation amortizes inference overhead across circuits.

\vspace{2mm}

\noindent\textbf{Execution flow.} Each step, the agent selects an action from the current mask. The environment applies the pass, updates the state, computes the reward, and checks termination. A \emph{skip} action advances to the next stage. If a pass requires follow-ups (e.g., retranslation after certain optimizations), the mask restricts selection until satisfied. Failed passes are masked from future selection in that episode. Passes with prerequisites (e.g., \texttt{VF2PostLayout} requires routing) remain masked until met. We wrap execution in a configurable timeout to handle passes that enter long-running loops; timeouts return a penalty and mark the pass as failed.

\vspace{2mm}

\noindent\textbf{Feature extraction.} The dual encoder computes DAG layer depths to bin gates by execution time. For post-routing stages, we extract noise characteristics from backend calibration: two-qubit error rates per edge and $T_1$/$T_2$ times per qubit, normalized by $500\mu s$ and clipped to $[0, 2]$. Global features capture topology statistics and circuit-backend compatibility ratios. We encode the last five passes and per-pass application counts to detect optimization convergence.

\vspace{2mm}

\noindent\textbf{RL approach.} We train with MaskablePPO~\cite{huang2020closer} from \textit{sb3-contrib}~\cite{raffin2021stable}. Proximal Policy Optimization (PPO)~\cite{schulman2017proximal} suits our formulation: its clipped surrogate objective provides stable updates despite variable episode lengths, and the maskable variant integrates with our action masking to assign probability mass only to valid passes. 

We evaluated alternatives within the same formulation (state representation, action masking, reward structure). DQN~\cite{mnih2013playing} struggled with cross-stage credit assignment under sparse terminal rewards: for example, on QPE and Grover, TVD improvement dropped to 5.8\% and -5\% respectively (vs.\ \sol{}'s 34\% and 16\%, over Qiskit (Fidelity Optimized)). A2C~\cite{mnih2016asynchronous} exhibited high episode-level variance from single-step updates, with TVD improvement on Deutsch-Jozsa falling to -66\%  (over Qiskit (Fidelity Optimized)), and exhibiting erratic compile-time behavior across benchmarks. SAC~\cite{haarnoja2018soft} required a continuous-to-discrete conversion that loses pass-selection precision. PPO with action masking avoids these issues and delivers consistent improvements across all benchmarks.

We train runs across 8 parallel workers with GPU acceleration. We apply soft normalization ($r' = r / \sqrt{1 + r^2}$) to prevent value function explosion from high-variance ESP ratios. We adopt standard PPO hyperparameters~\cite{andrychowicz2020matters}: learning rate $3 \times 10^{-4}$, 2048 steps per update, batch size 64, $\gamma = 0.99$, GAE~\cite{schulman2015high} $\lambda = 0.95$. For noise robustness (\req{2}), we train on 30 perturbed backend instances with Gaussian noise on error rates ($\sigma = 0.2\epsilon$) and coherence times ($\sigma = 0.1T$), while randomly disabling coupling edges (probability 0.05).

\vspace{2mm}

\noindent\textbf{Execution on real quantum backends.} We evaluate \methodname{} on currently operational IBM Quantum processors available through IBM Quantum Cloud: IBM Torino (133-qubit Heron r1), IBM Fez (156-qubit Heron r2), IBM Kingston (133-qubit Heron r2), and IBM Pittsburgh (156-qubit Heron r3). These processors feature tunable couplers that deliver 3--5$\times$ lower error rates compared to previous generations~\cite{ibmQuantumSystem}. All processors use heavy-hexagonal connectivity~\cite{chamberland2020topological}. \textit{This mix of architectures (across generations) and qubit counts tests whether \methodname{} generalizes across hardware variations}.

\begin{figure}[!t]
  \centering
  \includegraphics[width=\columnwidth]
  {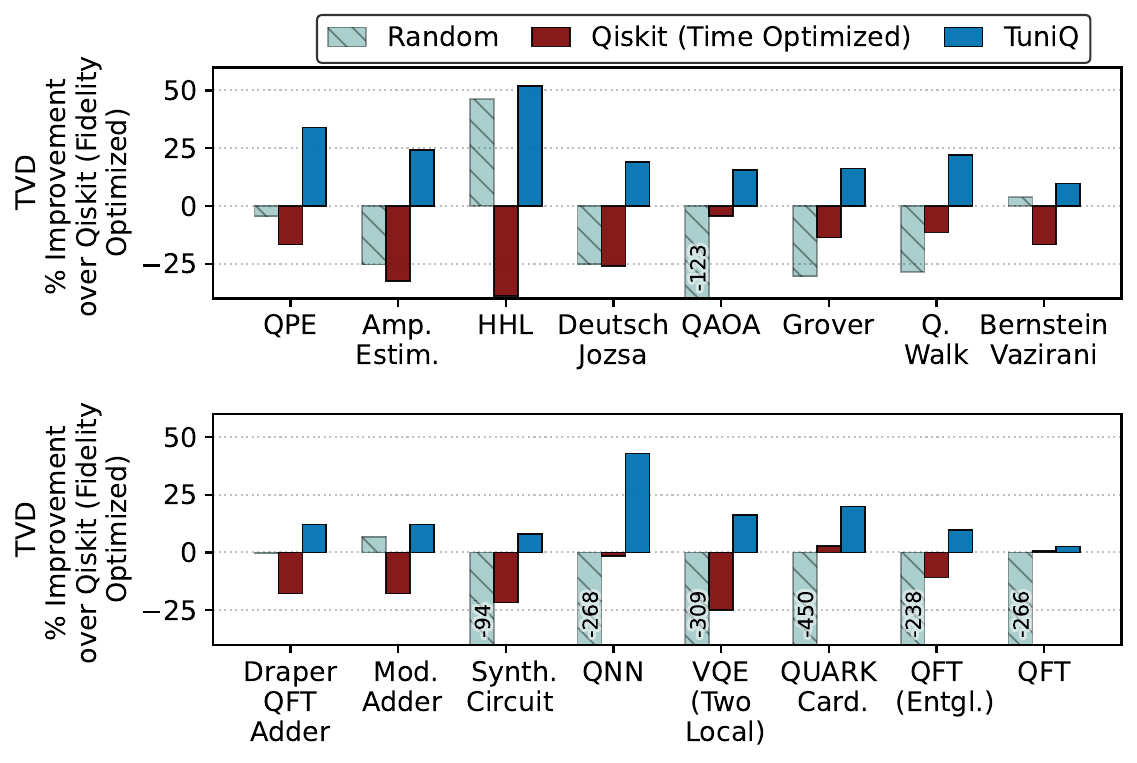}
  \vspace{-8mm}
  \caption{\methodname{} achieves higher fidelity (expressed as total variation distance (TVD)) compared to competitive methods.}
  \label{fig:result_tvd_different_circuits}
   \vspace{-6mm}
\end{figure}

\begin{figure}[!t]
  \centering
  \includegraphics[width=\columnwidth]{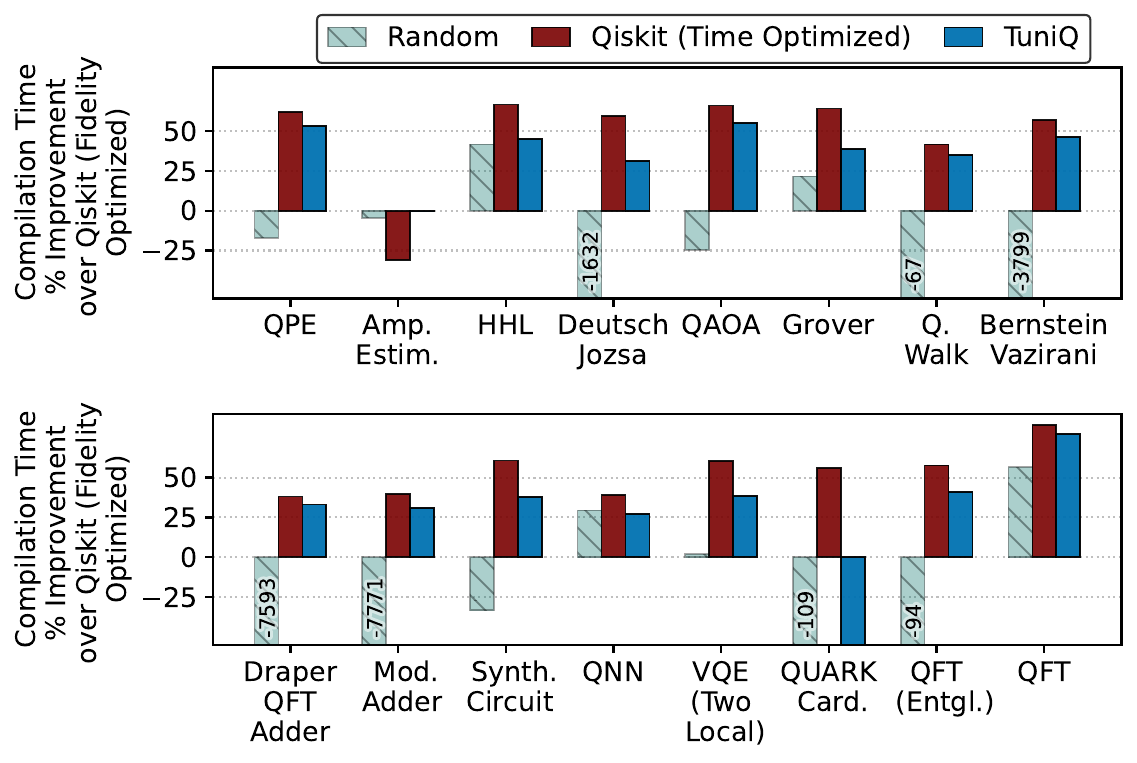}
  \vspace{-8mm}
  \caption{In most cases, \methodname{} results in less compilation time compared to the state-of-the-art solution.}
  \label{fig:results_time_different_circuits}
   \vspace{-6mm}
\end{figure}

\textcolor{black}{Our main effectiveness results (Figs.~\ref{fig:result_tvd_different_circuits},~\ref{fig:results_time_different_circuits}) and cross-backend generalization study (Fig.~\ref{fig:tvd_improvement_backends}) execute entirely on real IBM Quantum hardware via IBM Quantum Cloud. Controlled noise-scaling and decoherence-sensitivity studies (Figs.~\ref{fig:tvd_improvement_noises},~\ref{fig:t1_t2_noises}) use simulation, as these require systematic parameter sweeps that are infeasible on real hardware.} We extract calibration data through Qiskit Runtime Service, including per-qubit $T_1$/$T_2$ coherence times, gate error rates, and readout errors \textcolor{black}{-- to use in \sol{}'s noise-aware features, reward computation, and the controlled simulation studies}. Each circuit executes with 8192 shots for reliable TVD calculation. \textcolor{black}{Each data point is averaged across multiple runs across several calibration drifts.} We also perform a scaling study, with larger qubit versions of representative benchmarks. To evaluate robustness across noise conditions, we also use Qiskit Aer's~\cite{qiskit2024} density-matrix simulator with noise models constructed from the same IBM calibration data, scaling error rates and coherence times to simulate varying noise regimes. Simulations run on 8 NVIDIA H200 GPUs with 16384 shots per circuit -- a standard methodology in the quantum systems research community ~\cite{wang2025accelerating, suzuki2021qulacs}. Throughout our evaluation, we explicitly state which results are obtained via simulation.

\vspace{1mm}

\noindent\textbf{Benchmark circuits.} \textit{We evaluate on two circuit categories.} For algorithmic benchmarks, we use MQTBench~\cite{quetschlich2023mqt}, a widely adopted benchmark suite in quantum systems and compiler research ~\cite{nguyen2024qfaas, li2024quarl, quetschlich2023compiler}. We select circuits at the algorithmic abstraction level spanning 5--15 qubits for hardware evaluation. These benchmarks cover a diverse range of circuit characteristics: varying entanglement patterns from local nearest-neighbor to all-to-all connectivity, gate compositions ranging from Clifford-dominated to rotation-heavy circuits, and circuit structures from highly regular (repeated subroutines) to irregular problem-dependent ansätze. This diversity ensures our evaluation captures the range of compilation challenges encountered in practice. To assess scaling, we evaluate larger variants up to 65 qubits; in the quantum context, circuits beyond 40 qubits approach utility-scale workloads that exceed classical simulability thresholds and stress-test compilation on near-full-device mappings ~\cite{kim2023evidence, arute2019quantum}. For generalization testing, we use Qiskit's random circuit generator with qubit counts of 5--15 and depths $2n$ to $5n$ where $n$ is qubit count. \textit{Training uses random circuits exclusively}; evaluation on algorithmic benchmarks tests if learned policies transfer to structured circuits unseen during training (\req{1} from Sec.~\ref{sec:des_req}).

\vspace{2mm}

\noindent\textbf{Competing solutions.} Our primary baseline is \emph{Qiskit (Fidelity Optimized)}, which corresponds to Level 3, the most aggressive configuration and current state-of-the-art; it enables all available optimization passes and serves as our primary baseline. We also compare with \emph{Qiskit (Time Optimized)}, which corresponds to Level 0, performing minimal compilation without optimization~\cite{qiskit2024}. Beyond fixed strategies, we evaluate \emph{random selection} (uniform sampling from valid actions), \emph{greedy selection} (maximize immediate ESP at each stage), and \emph{Evolutionary optimization} (Covariance Matrix Adaptation Evolution Strategy 
(CMA-ES)~\cite{akiba2019optuna, hansen2001completely}. These alternatives test whether simpler approaches suffice or whether RL's sequential, reward-shaped learning provides fundamental advantages.

\vspace{2mm}

\noindent\textbf{Evaluation metrics.} We report total variation distance (TVD) as the primary metric for fidelity and compilation time (includes inference time to determine passes). They are expressed as a \% improvement over Qiskit (Fidelity Optimized) (greater is better).

\section{Evaluation}
\label{sec:eval}

Here, we show the overall performance of \sol{}. Then we analyze the reasons behind its effectiveness and how the performance scales.

\subsection{Effectiveness  of \sol{}}
\noindent\textbf{Fidelity improvement.} As shown in Fig.~\ref{fig:result_tvd_different_circuits}, \methodname{} improves TVD on all representative benchamrks from MQTBench (Sec.~\ref{sec:implementation}), with an average gain of 20\%. In the NISQ regime, even modest TVD reductions have an outsized impact: errors compound exponentially with circuit depth. For example, for QPE, \methodname{} reduces TVD (raw numbers) from 0.76 to 0.50; at 0.76, the output distribution is dominated by incorrect states and the algorithm effectively fails, while at 0.50 the correct eigenvalue emerges as the plurality outcome. For variational algorithms like VQE and QAOA, lower TVD translates directly to more accurate energy estimates, which reduces the iterations for convergence and improves optimization quality.

The improvements hold across structured algorithms (QPE, Grover, Deutsch-Jozsa) and variational workloads (VQE, QNN, QAOA), despite \methodname{} training exclusively on random circuits. This transfer demonstrates that the policy learns generalizable circuit features rather than memorizing specific structures. Qiskit (Time Optimized) and random selection degrade fidelity significantly, which confirms that gains stem from learned decisions, instead of action space design. One exception is that in HHL, random selection achieves 46\% improvement, close to \methodname{}'s 52\%. HHL's phase rotations are disrupted by aggressive optimization, so even random avoidance of certain passes outperforms Qiskit (Fidelity Optimized) solution. \methodname{} identifies this pattern systematically. \textcolor{black}{All reported TVD improvements are averaged across multiple calibration windows to capture hardware drift; \sol{}'s inference is deterministic given a fixed calibration, so observed variation (standard deviation of less than 5\%) reflects calibration drift rather than policy stochasticity. }

\vspace{1mm}

\noindent\textbf{Compilation time.} \methodname{} reduces compilation time by an average of 34\%. (Fig.~\ref{fig:results_time_different_circuits}). For variational algorithms like VQE and QAOA, this improvement compounds: a single optimization run may recompile the same circuit thousands of times with different parameters, so a modest reduction in compilation time translates to hours saved in end-to-end execution. The agent achieves these reductions by learning which passes contribute to fidelity and skipping redundant transformations; Qiskit (Fidelity Optimized) applies its full pass sequence uniformly regardless of circuit structure. 

One case deviates: on QUARK Cardinality, \methodname{} incurs longer compilation as it selects \textit{LookaheadSwap} pass for routing. This pass explores more SWAP candidates, increasing compilation cost but producing fewer two-qubit gates and 20\% better TVD. \sol{}'s agent has learned through routing explorations that the overhead is justified. The random baseline shows extreme variance, having significant overhead on some benchmarks due to pathological pass interactions, which highlights the cost of uninformed selection. Qiskit (Time Optimized) compiles faster, but at a severe fidelity cost, as it applies only minimal passes to satisfy hardware constraints. It serves as a time-only floor rather than a quality-competitive baseline: compared to Qiskit (Time Optimized), \sol{} incurs 47\% additional compilation time while improving TVD by 30\%; compared to Qiskit (Fidelity Optimized), \sol{} improves TVD by 20\% and reduces compilation time by 34\%. Our goal is a better quality--time tradeoff than the strongest quality baseline, not to undercut the time-only floor. \textcolor{black}{Compilation-time results are likewise averaged across calibration windows (standard deviation of 4.6\%)}


\begin{table}[t]
\centering
\caption{{\sol{}'s improvements on QASMBench circuits \textnormal{(expressed as \%-improvement over Qiskit (Fidelity Optimized))}.}}
\label{tab:qasmbench}
\vspace{-3mm}
{\footnotesize
\setlength{\tabcolsep}{3pt}
\renewcommand{\arraystretch}{1.05}
\begin{tabular}{@{}lcc@{}}
\toprule
\textbf{Benchmark} &
\textbf{\shortstack{TVD\\(\% Improvement)}} &
\textbf{\shortstack{Compilation Time\\(\% Improvement)}} \\
\midrule
QPE      & 39.0 & 38.6 \\
QFT      & 12.0 & 38.3 \\
QNN      & 3.8  & 22.4 \\
QWALK    & 6.0  & 51.0 \\
TOFFOLI  & 7.0  & 42.0 \\
WSTATE   & 10.0 & 44.0 \\
QAOA     & 27.0 & 49.0 \\
VQE      & 14.3 & 31.5 \\
\bottomrule
\end{tabular}}
\vspace{-4mm}
\end{table}

\vspace{1mm}

\noindent\textbf{Cross-benchmark validation on QASMBench.} To validate that \sol{}'s gains are not specific to MQTBench, we evaluate on QASMBench circuits spanning structured algorithms (QPE, QFT, Toffoli), variational workloads (VQE, QAOA, QNN), and state preparation routines (WSTATE, QWALK). Table~\ref{tab:qasmbench} reports improvements over Qiskit (Fidelity Optimized). \sol{} delivers consistent positive improvements across all eight circuits: TVD improves by 3.8--39\% and compile time reduces by 22--51\%, in line with the gains observed on MQTBench. This consistency across two distinct benchmark suites supports that the policy generalizes to algorithmic structure, and not to the specifics of any single benchmark distribution.

\subsection{Robustness Across Hardware and Noise}
\label{sec:cross_backend}

A practical compiler should maintain performance as hardware conditions vary. We evaluate \methodname{} along three axes: across backends with different topologies and calibrations, under scaled noise levels simulating future hardware improvements, and under varying coherence times. For brevity, we present results for a subset of benchmarks that have the most diversity in circuit characteristics: deep controlled-unitary structures, oracle-based circuits, amplitude amplification patterns, and variational ansätze with dense entanglement. Similar trends hold across the remaining benchmarks.

\vspace{1mm}

\begin{figure}[!t]
  \centering
  \includegraphics[width=\columnwidth]{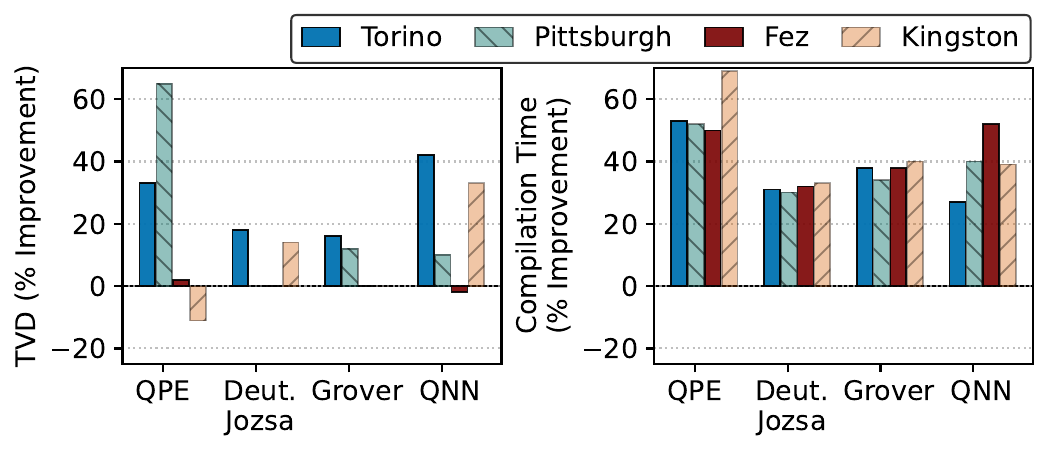}
  \vspace{-9mm}
  \caption{\sol{} improves fidelity and compile time across different backends, even when trained on a fixed backend \textnormal{(expressed as \%-improvement over Qiskit (Fidelity Optimized))}.}
  \label{fig:tvd_improvement_backends}
  \vspace{-2mm}
\end{figure}

\noindent\textbf{Cross-backend generalization.} Fig.~\ref{fig:tvd_improvement_backends} reports results across four IBM backends: Torino (Heron R1, 133 qubits), Fez (Heron R2, 156 qubits), Kingston (Heron R2, 133 qubits), and Pittsburgh (Heron R3, 156 qubits). \methodname{} trains on Pittsburgh with perturbed connectivity and noise. Evaluation on the other three backends is zero-shot, with no retraining. The policy transfers successfully: TVD improves on 14 of 16 benchmark-backend pairs, with gains up to 64\% (QPE on Pittsburgh) while compilation time reduces by 27--67\% consistently. Two cases show slight degradation: QPE on Kingston and QNN on Fez, both Heron R2 devices. R2 processors have different calibration characteristics than R3, and the policy trained on R3 encounters out-of-distribution noise patterns; deep circuits like QPE and heavily entangled circuits like QNN amplify sensitivity to these differences.

\vspace{1mm}

\begin{figure}[!t]
  \centering
  \includegraphics[width=\columnwidth]{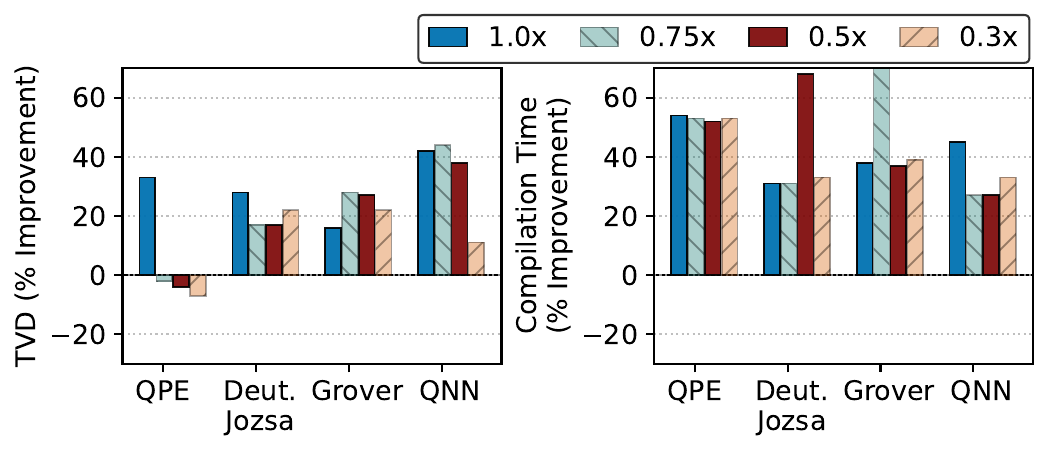}
  \vspace{-8mm}
  \caption{\sol{} is effective across hardware noise levels \textnormal{(expressed as \%-improvement over Qiskit (Fidelity Optimized)). \textcolor{black}{This is a simulation-based controlled study.}}}
  \label{fig:tvd_improvement_noises}
  \vspace{-4mm}
\end{figure}

\noindent\textbf{Noise scaling.} 
\textit{As a controlled sensitivity study to characterize how \sol{}'s gains evolve as hardware improves}, we scale gate and readout errors to 0.75$\times$, 0.5$\times$, and 0.3$\times$ of current calibration values (scaling factors chosen to reflect error-rate reductions observed across IBM device generations). Scaling is applied to each qubit's and each gate's individual calibrated error rate from the IBM snapshot, preserving device-specific heterogeneity rather than collapsing to a global average. We focus on gate and readout errors and $T_1$/$T_2$ as the dominant calibration terms exposed through Qiskit Runtime; this is a sensitivity study, not a full hardware emulator, and omits effects such as crosstalk and gate-type-specific error variation. Fig.~\ref{fig:tvd_improvement_noises} shows \methodname{} maintains positive TVD improvement across most benchmarks and noise levels, with compilation time benefits persisting at 27--70\% (on IBM Torino). As noise decreases, the relative advantage over Qiskit (Fidelity Optimized) narrows for some circuits: at 0.3$\times$ noise, QPE shows slight degradation. This is expected -- when hardware errors are low, the marginal benefit of optimized pass selection diminishes since even suboptimal circuits execute with reasonable fidelity. The compilation time advantage remains valuable regardless, showing the effectiveness of \sol{}.

\vspace{1mm}

\begin{figure}[!t]
  \centering
  \includegraphics[width=\columnwidth]{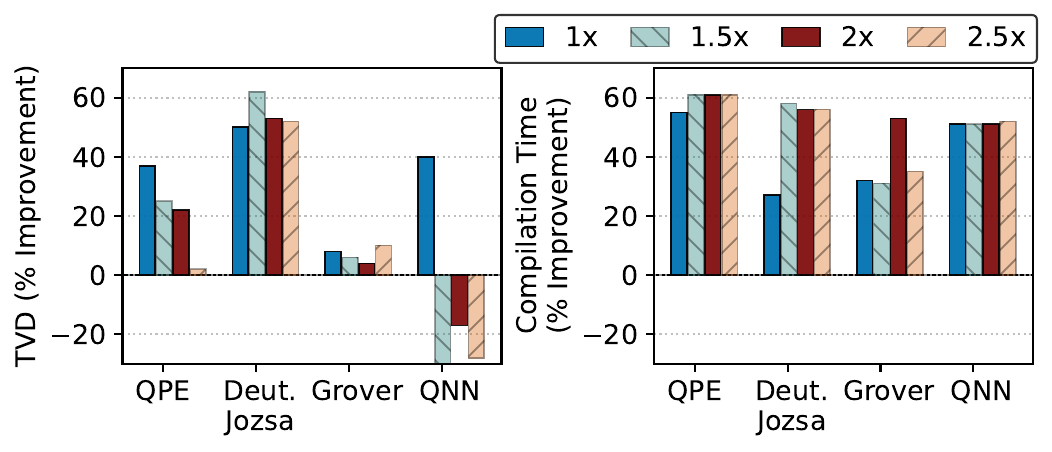}
  \vspace{-8mm}
  \caption{\sol{} is effective for different decoherence times \textnormal{(expressed as \%-improvement over Qiskit (Fidelity Optimized)). \textcolor{black}{This is a simulation-based controlled study.}}}
  \label{fig:t1_t2_noises}
  \vspace{-2mm}
\end{figure}

\noindent\textbf{Decoherence time scaling.} We scale $T_1$/$T_2$ to 1.5--2.5$\times$ baseline to simulate longer coherence, disabling gate errors to isolate decoherence effects. This decoupling is a deliberate ablation, not a claim that gate errors and decoherence are separable on live hardware; the goal is to probe the policy's sensitivity to each error mechanism in isolation. Fig.~\ref{fig:t1_t2_noises} shows \methodname{} generalizes across coherence regimes, with Deutsch-Jozsa showing 50--60\% TVD improvement. QNN degrades at 2.5$\times$: with gate errors removed, noise-aware routing (trained to avoid high-error edges) becomes counterproductive by selecting longer paths that add unnecessary depth. This is informative rather than a failure mode -- it confirms the policy is genuinely sensitive to the balance between edge error rates and depth/decoherence, which is the intended behavior for a noise-aware policy. As hardware evolves, periodic retraining can improve gains. Results in Figures~\ref{fig:tvd_improvement_noises} and~\ref{fig:t1_t2_noises} are derived via simulation.


\subsection{Why \sol{} Works?} 
\label{sec:ss}


\begin{table}[t]
\centering
\caption{Contribution of each design component (QPE benchmark, similar trends for others) \textnormal{(expressed as \%-improvement over Qiskit (Fidelity Optimized))}.}
\label{tab:ablation_method}
\vspace{-3mm}
{\footnotesize
\setlength{\tabcolsep}{3pt}
\renewcommand{\arraystretch}{1.05}
\begin{tabular}{@{}lcc@{}}
\toprule
\textbf{Method} &
\textbf{\shortstack{TVD\\(\% Improvement)}} &
\textbf{\shortstack{Compilation Time\\(\% Improvement)}} \\
\midrule
\textbf{TuniQ (Full)}         & 34.0 & 53.3 \\
Stage-RL (Init+Opt)           & 16.7 & -72.8 \\
Reward: Gate+Depth            & 27.4 & 50.7 \\
Features: NoNoise             & 34.0 & 46.9 \\
Features: 2Q-only             & 31.1 & 47.5 \\
\bottomrule
\end{tabular}}
\vspace{-4mm}
\end{table}

\noindent\textbf{Component Analysis.} Table~\ref{tab:ablation_method} addresses the natural design questions that arise from our formulation. The first question is whether RL control over all stages is necessary, or whether layout and routing can use Qiskit defaults while RL handles only init and optimization (which has the most number of passes affecting the circuit depth and number of gates). \emph{Stage-RL (Init+Opt)} tests this: TVD improvement drops from 34\% to 17\%, and compilation time degrades by 73\%. The reason is that layout and routing choices propagate through the pipeline, and optimizing only the surrounding stages cannot compensate -- they need to be jointly considered.

The second question is whether a complex ESP-based reward is necessary, or whether simpler gate count and depth metrics suffice. \emph{Reward: Gate+Depth} shows TVD drops to 27\%: hardware-aware fidelity signals matter because gate count alone ignores error rate variations across qubit pairs. The third question is whether noise-augmented features justify their complexity. \emph{Features: NoNoise} shows TVD holds, but compilation time improvement drops from 53\% to 47\%; without noise information, the agent cannot recognize when aggressive optimization is unnecessary and applies redundant passes, confirming the importance of real-time calibration data in the state representation. Finally, \emph{Features: 2Q-only} confirms that single-qubit features provide a modest but measurable signal, though two-qubit interactions dominate.

\vspace{2mm}

\begin{figure}[!t]
  \centering
  \includegraphics[width=\columnwidth]{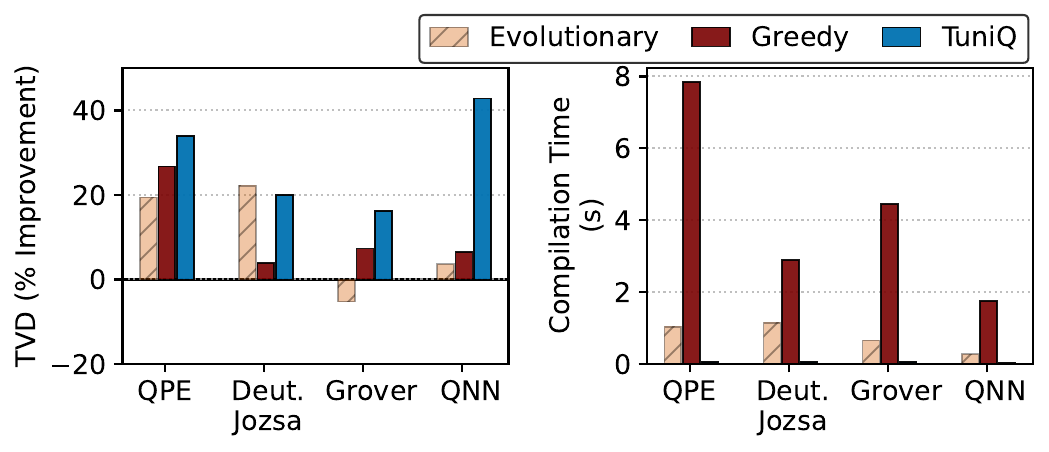}
  \vspace{-8mm}
  \caption{\sol{}'s is more effective than other searches \textnormal{(expressed as \%-improvement over Qiskit (Fidelity Optimized))}.}
  \label{fig:bbo_greedy} 
  \vspace{-4mm}
\end{figure}

\noindent\textbf{Pass-selection patterns.} To examine whether \sol{} learns recurring algorithmic structure, we record the frequency with which each pass is selected across benchmarks. Selections are strongly circuit-dependent: CommutativeCancellation is selected on 100\% of HHL episodes, but only 33\% of Grover episodes, reflecting the different roles gate cancellation plays for each algorithm. Several passes that Qiskit (Fidelity Optimized) applies uniformly are consistently skipped by \sol{} -- e.g., InverseCancellation and ElidePermutations. This indicates that an always-on application provides no benefit and that selective skipping contributes to compile-time gains. Selections are also backend-dependent: for Grover, ContractIdleWiresInControlFlow is selected 67\% of the time on Pittsburgh but 0\% on Fez and Kingston, reflecting topology and calibration differences. Together, these patterns confirm that simply enabling all available passes is not optimal, and that \sol{}'s gains come from recognizing when passes help and when they hurt.

\vspace{2mm}

\noindent\textbf{Search-based alternatives.} Beyond ablating components, we ask whether RL is necessary at all. Pass selection is discrete and non-differentiable, which makes gradient-based methods hard to apply. The configuration space is combinatorial and lacks smoothness assumptions for search techniques like Bayesian optimization. The natural alternatives are evolutionary search and greedy heuristics.

We implement two baselines over the same pass space. \emph{Greedy} samples $N$ valid passes at each stage, applies each, and selects the one with the best gate count and depth reduction. \emph{Evolutionary} uses Optuna's CMA-ES (Covariance Matrix Adaptation Evolution Strategy) to evolve complete pass configurations as vectors, using ESP as fitness. CMA-ES adapts its search covariance based on successful candidates, making it effective for black-box optimization without gradients. We give Evolutionary a generous time budget to ensure sufficient iterations to converge.

Fig.~\ref{fig:bbo_greedy} shows \methodname{} achieves the best TVD on three of four benchmarks. Evolutionary slightly outperforms on Deutsch-Jozsa (22\% vs 20\%), but the comparison requires context: Greedy and Evolutionary incur per-circuit search overhead (1--8 seconds), while \methodname{}'s training is one-time and only inference appears in compilation (under 0.1 seconds). For variational workloads recompiling thousands of times, this compounds to hours saved. Evolutionary also fails on Grover ($-$5\%) because without learned representations, search converges to local optima that score well on fitness but produce suboptimal circuits. This evolutionary baseline corresponds to the direction \'Swierkowska et al.~\cite{swierkowska2024achieving} suggested as future work to mitigate their NSGA-II search latency (4--9 seconds per circuit). Our experiment confirms the gap: per-circuit search consumes 1--8 seconds with 28.3\% worse TVD on average, because each new circuit requires fresh search iterations and variational workloads recompile thousands of times. This motivated our shift to an RL-based approach that amortizes search to a single forward pass.

\subsection{Performance Scaling}

\begin{figure}[!t]
  \centering
  \includegraphics[width=\columnwidth]{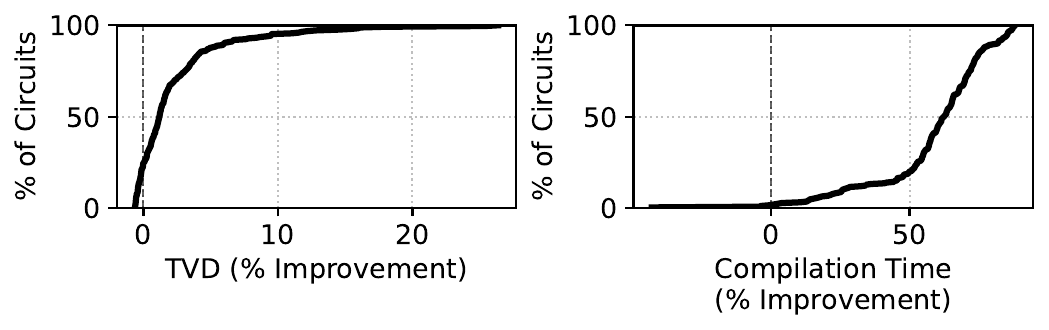}
  \vspace{-8mm}
  \caption{\sol{}'s improvement on randomly generated circuits \textnormal{(\%-improvement over Qiskit (Fidelity Optimized))}.}
  \label{fig:cdf_random_tvd_time_improvement} 
   \vspace{-4mm}
\end{figure} 

\noindent\textbf{Random circuits.} Algorithmic benchmarks test specific circuit families, but a practical compiler should handle arbitrary circuits. We evaluate on 1000 random circuits (5--15 qubits, depths 10--50), distinct from the training set. Fig.~\ref{fig:cdf_random_tvd_time_improvement} shows cumulative distributions over Qiskit (Fidelity Optimized). Compilation time improves consistently, with 80\% of circuits showing 50--70\% reduction. TVD improvement is modest but reliably positive: median improvement of 10\%, with occasional gains exceeding 20\%. Near-zero degradation cases are rare, indicating \methodname{} does not sacrifice fidelity for speed by learning when to skip redundant passes.

\vspace{2mm}

\begin{figure}[!t]
  \centering
  \includegraphics[width=\columnwidth]{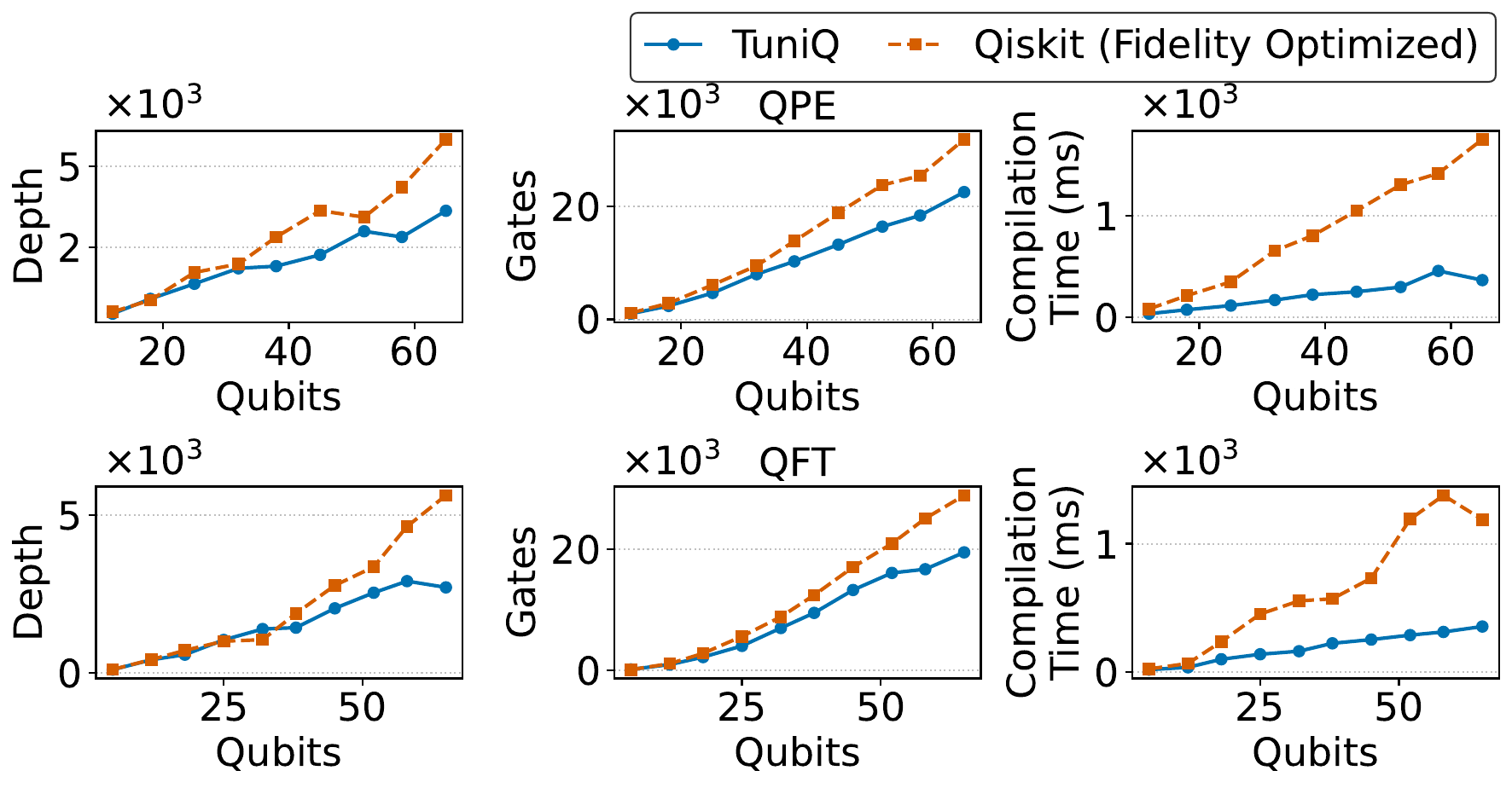}
  \vspace{-8mm}
  \caption{\methodname{} helps in strong scaling -- passes learned on small circuits transfer to larger ones with growing advantage.}
   \vspace{-4mm}
  \label{fig:record_replay_scaling} 
  
\end{figure}

\noindent\textbf{Strong scaling.} Beyond generalizing to unseen circuits, a practical compiler should scale with circuit size. The same algorithm can be instantiated at different qubit counts depending on problem size; retraining for each scale would be prohibitive. Weak scaling (batch throughput) is orthogonal to per-circuit quality: compilation runs user-side before circuits are submitted to the quantum cloud, and \sol{}'s inference is independent across circuits, so batch size affects throughput but not the quality of any single compilation. \textcolor{black}{This experiment is deliberately constrained: rather than running per-circuit inference (\sol{}'s deployment mode), we freeze the pass sequence learned on a small instance and replay it at larger scales. This tests whether the policy captures recurring algorithmic structure, providing a conservative lower bound on \sol{}'s scaling behavior.} To test strong scaling, we infer pass selections on small instances (5 qubits for QFT, 12 qubits for QPE) and apply the same sequence to larger instances up to 65 qubits. \textcolor{black}{In deployment, \sol{} performs per circuit inference at every scale; the results here represent a stricter setting where the agent does not see the larger circuit at all.} Fig.~\ref{fig:record_replay_scaling} reports depth, gate count, and compilation time, instead of the TVD: executing 60+ qubit circuits on current hardware is prohibitively expensive, and NISQ devices produce too much noise for meaningful fidelity measurements at this scale. Depth and gate count are appropriate proxies because each additional gate introduces error, and deeper circuits suffer more decoherence. Reducing metrics like circuit depth and number of gates is how compilers like Qiskit improve circuit fidelity.


\methodname{}'s advantage grows with circuit size. At 65 qubits, circuits have 40\% fewer gates and 50\% lower depth than Qiskit (Fidelity Optimized), with 2--3$\times$ faster compilation. The gap widens because the policy captures algorithmic structure rather than circuit-specific features: QFT's recursive pattern and QPE's controlled-unitary ladder remain consistent across scales. \textit{Even in this constrained transfer setting where no per-circuit inference is performed at scale, depth, gate count, and compile time, all improve over Qiskit (Fidelity Optimized) as circuits grow -- encouraging for HPC workflows that integrate quantum subroutines as problem sizes scale toward utility.}


\vspace{2mm}

\noindent\textbf{Per-circuit RL inference at scale.} Beyond the transferred-sequence experiment above, we additionally ran per-circuit RL inference on QFT and QPE at 30--50 qubits to verify deployment-mode performance directly at scale. \sol{} improves compilation time by 68\%, gate count by 27\%, and depth by 25\% over Qiskit (Fidelity Optimized). Search-based alternatives are infeasible at this scale: each candidate configuration requires a full compilation, and the search budget that produced the 1--8 seconds of overheads in Sec.~\ref{sec:ss} grows prohibitively at 30--50 qubits. Environmental noise beyond standard calibration is not captured by any current compiler, so the effect of it is shared across all the evaluated baselines.


  


\section{Related Works} 
\label{sec:related_work}

\noindent\textbf{Quantum Circuit Compilation and Stage-Specific Optimization.}
Quantum circuit compilation transforms abstract circuits into hardware-executable instructions through a multi-stage pipeline~\cite{qiskit2024}. A significant amount of work has targeted individual stages of this pipeline to improve execution fidelity on noisy hardware ~\cite{murali2019noise, tannu2019not, li2019tackling}. Nam et al.~\cite{nam2018} developed automated circuit rewriting rules, while Patel et al.~\cite{patel2022charter} showed that identifying the most critical gate operations via amplified reversibility enables targeted optimization of the gates that matter most. Reinforcement learning has also been applied to routing~\cite{quetschlich2023compiler}, demonstrating that learned policies can outperform hand-crafted heuristics. Ravi et al.~\cite{ravi2023navigating} characterized how quantum hardware noise varies temporally and across devices, emphasizing the need for continual noise-aware transpilation. Similarly, Huo et al.~\cite{huo2025revisiting} showed that aggressive optimization levels often yield negligible fidelity gains over lighter configurations, reinforcing that indiscriminate pass application is wasteful. 
\textit{However, these works optimize individual compilation stages in isolation. No study has shown how pass selections interact across stages: a locally optimal pass can inflate gate counts in ways that later optimization cannot recover, and an aggressive init-stage simplification may eliminate structure that layout algorithms exploit.} 

\vspace{2mm}

\noindent\textbf{Quantum computing as an HPC accelerator.}
Quantum processors are increasingly deployed as heterogeneous accelerators within HPC ecosystems~\cite{beck2024integrating, 9537178, alexeev2024quantum, chundury2024qfw,chundury2025quantum,
mittal2025openmp,zhang2023disq, humble2021quantum}. IBM's quantum-centric supercomputing vision~\cite{bravyi2022future}, NVIDIA's CUDA-Q~\cite{nvidiaNVIDIACUDAQ}, and cloud orchestration~\cite{giortamis2025qonductor} dispatch circuits from classical workflows. HPC-scale simulation advances include GPU acceleration~\cite{patti2025augmenting}, hierarchical partitioning~\cite{xu2024atlas}, tensor networks~\cite{fu2024surpassing}, and compression~\cite{zhang2025bmqsim}. Dominant near-term applications are variational algorithms~\cite{peruzzo2014variational, farhi2014quantum, havlivcek2019supervised, cerezo2021variational, bharti2022noisy} and NISQ workloads in ML~\cite{biamonte2017quantum, silver2024lexiql} and optimization~\cite{shaydulin2024evidence}. These workloads are compilation-sensitive: suboptimal transpilation inflates gates, degrades gradients, and induces barren plateaus~\cite{wang2021noise}. Mitigations include calibration-aware transpilation~\cite{9951287} and noise-aware scheduling~\cite{huo2025anchor}, yet most treat transpilation as a black box approach. Closer to our setting, \'Swierkowska et al.~\cite{swierkowska2024achieving} apply per-circuit NSGA-II design-space exploration over LLVM/QIR pass subsets and sequences in the Munich Quantum Compiler, optimizing structural metrics (gates, depth, entanglement, parallelism) over a different pass ecosystem. \sol{} instead selects passes across all stages of Qiskit's transpilation pipeline (init, layout, routing, translation, optimization, cleanup) and conditions on hardware noise. \textit{To the best of our knowledge, \sol{} is the first noise-conditioned system that jointly optimizes pass selection across the full IBM Qiskit circuit transpilation pipeline.}

\vspace{-2mm}
\section{Conclusion} 
\label{sec:conclusion}

\methodname{} is the first adaptive pass selector for quantum compilation, which replaces the fixed pass application sequence used by state-of-the-art compilers, with learned circuit- and noise-aware selection across the entire transpilation pipeline stages. It improves fidelity, reduces compile time, generalizes across backends without retraining, and scales to utility-scale circuits with growing advantage as quantum and classical HPC system integration matures. \sol{} is \textit{open-sourced} at: \hyperlink{https://zenodo.org/records/19969999}{https://zenodo.org/records/19969999}. We hope that this work helps the community to build adaptive compilation pass selection frameworks as hardware and transpilers mature.




\vspace{2mm}

\noindent{\textbf{Acknowledgments.} We thank the reviewers for their constructive feedback. IBM Quantum resources were used for this work. All positions expressed in the paper are those of the authors and do not reflect the position of the IBM Quantum team. This research used the resources of the National Energy Research Scientific Computing Center, a DOE Office of Science User Facility supported by the Office of Science of the U.S. Department of Energy under Contract No. DE-AC02-05CH11231 using NERSC award NERSC DDR-ERCAP0037923. This work is supported by University of Utah’s Kahlert School of Computing, Price College of Engineering, Scientific Computing \& Imaging (SCI) Institute, and Rice University.}
\bibliographystyle{ACM-Reference-Format}
\balance
\bibliography{references}

@Article{Nam2018,
author={Nam, Yunseong
and Ross, Neil J.
and Su, Yuan
and Childs, Andrew M.
and Maslov, Dmitri},
title={Automated optimization of large quantum circuits with continuous parameters},
journal={npj Quantum Information},
year={2018},
month={May},
day={10},
volume={4},
number={1},
pages={23},
issn={2056-6387},
doi={10.1038/s41534-018-0072-4},
url={https://doi.org/10.1038/s41534-018-0072-4}
}

@inproceedings{quetschlich2023compiler,
  title={Compiler optimization for quantum computing using reinforcement learning},
  author={Quetschlich, Nils and Burgholzer, Lukas and Wille, Robert},
  booktitle={2023 60th ACM/IEEE Design Automation Conference (DAC)},
  pages={1--6},
  year={2023},
  organization={IEEE}
}

@inproceedings{ravi2023navigating,
  title={Navigating the dynamic noise landscape of variational quantum algorithms with qismet},
  author={Ravi, Gokul Subramanian and Smith, Kaitlin and Baker, Jonathan M and Kannan, Tejas and Earnest, Nathan and Javadi-Abhari, Ali and Hoffmann, Henry and Chong, Frederic T},
  booktitle={Proceedings of the 28th ACM International Conference on Architectural Support for Programming Languages and Operating Systems, Volume 2},
  pages={515--529},
  year={2023}
}

@article{huo2025revisiting,
  title={Revisiting Noise-adaptive Transpilation in Quantum Computing: How Much Impact Does it Have?},
  author={Huo, Yuqian and Wei, Jinbiao and Kverne, Christopher and Akewar, Mayur and Bhimani, Janki and Patel, Tirthak},
  journal={arXiv preprint arXiv:2507.01195},
  year={2025}
}

@inproceedings{xu2025optimizing,
  title={Optimizing quantum circuits, fast and slow},
  author={Xu, Amanda and Molavi, Abtin and Tannu, Swamit and Albarghouthi, Aws},
  booktitle={Proceedings of the 30th ACM International Conference on Architectural Support for Programming Languages and Operating Systems, Volume 1},
  pages={777--793},
  year={2025}
}

@ARTICLE{9537178,
  author={Humble, Travis S. and McCaskey, Alexander and Lyakh, Dmitry I. and Gowrishankar, Meenambika and Frisch, Albert and Monz, Thomas},
  journal={IEEE Micro}, 
  title={Quantum Computers for High-Performance Computing}, 
  year={2021},
  volume={41},
  number={5},
  pages={15-23},
  doi={10.1109/MM.2021.3099140}}

@article{bravyi2022future,
  title={The future of quantum computing with superconducting qubits},
  author={Bravyi, Sergey and Dial, Oliver and Gambetta, Jay M and Gil, Dar{\'\i}o and Nazario, Zaira},
  journal={Journal of Applied Physics},
  volume={132},
  number={16},
  year={2022},
  publisher={AIP Publishing}
}

@misc{nvidiaNVIDIACUDAQ,
	author = {},
	title = {{N}{V}{I}{D}{I}{A} {C}{U}{D}{A}-{Q} --- developer.nvidia.com},
	howpublished = {\url{https://developer.nvidia.com/cuda-q}},
	year = {},
}

@inproceedings{giortamis2025qonductor,
  title={Qonductor: A Cloud Orchestrator for Quantum Computing},
  author={Giortamis, Emmanouil and Romao, Francisco and Tornow, Nathaniel and Lugovoy, Dmitry and Bhatotia, Pramod},
  booktitle={Proceedings of the International Conference for High Performance Computing, Networking, Storage and Analysis},
  pages={728--745},
  year={2025}
}

@inproceedings{patti2025augmenting,
  title={Augmenting Simulated Noisy Quantum Data Collection by Orders of Magnitude Using Pre-Trajectory Sampling with Batched Execution},
  author={Patti, Taylor Lee and Nguyen, Thien and Lietz, Justin Gage and McCaskey, Alex J and Khailany, Brucek},
  booktitle={Proceedings of the International Conference for High Performance Computing, Networking, Storage and Analysis},
  pages={762--773},
  year={2025}
}

@article{beck2024integrating,
  title={Integrating quantum computing resources into scientific hpc ecosystems},
  author={Beck, Thomas and Baroni, Alessandro and Bennink, Ryan and Buchs, Gilles and P{\'e}rez, Eduardo Antonio Coello and Eisenbach, Markus and da Silva, Rafael Ferreira and Meena, Muralikrishnan Gopalakrishnan and Gottiparthi, Kalyan and Groszkowski, Peter and others},
  journal={Future Generation Computer Systems},
  volume={161},
  pages={11--25},
  year={2024},
  publisher={Elsevier}
}

@inproceedings{anagolum2024elivagar,
  title={{\'E}liv{\'a}gar: Efficient quantum circuit search for classification},
  author={Anagolum, Sashwat and Alavisamani, Narges and Das, Poulami and Qureshi, Moinuddin and Shi, Yunong},
  booktitle={Proceedings of the 29th ACM International Conference on Architectural Support for Programming Languages and Operating Systems, Volume 2},
  pages={336--353},
  year={2024}
}

@inproceedings{xu2024atlas,
  title={Atlas: Hierarchical partitioning for quantum circuit simulation on gpus},
  author={Xu, Mingkuan and Cao, Shiyi and Miao, Xupeng and Acar, Umut A and Jia, Zhihao},
  booktitle={SC24: International Conference for High Performance Computing, Networking, Storage and Analysis},
  pages={1--17},
  year={2024},
  organization={IEEE}
}

@inproceedings{fu2024surpassing,
  title={Surpassing Sycamore: Achieving Energetic Superiority Through System-Level Circuit Simulation},
  author={Fu, Rong and Su, Zhongling and Zhong, Han-Sen and Zhao, Xiti and Zhang, Jianyang and Pan, Feng and Zhang, Pan and Zhao, Xianhe and Chen, Ming-Cheng and Lu, Chao-Yang and others},
  booktitle={SC24: International Conference for High Performance Computing, Networking, Storage and Analysis},
  pages={1--20},
  year={2024},
  organization={IEEE}
}

@inproceedings{zhang2025bmqsim,
  title={BMQSim: Overcoming Memory Constraints in Quantum Circuit Simulation with a High-Fidelity Compression Framework},
  author={Zhang, Boyuan and Fang, Bo and Ye, Fanjiang and Guo, Luanzheng and Song, Fengguang and Tallent, Nathan and Tao, Dingwen},
  booktitle={Proceedings of the 39th ACM International Conference on Supercomputing},
  pages={689--704},
  year={2025}
}

@article{shaydulin2024evidence,
  title={Evidence of scaling advantage for the quantum approximate optimization algorithm on a classically intractable problem},
  author={Shaydulin, Ruslan and Li, Changhao and Chakrabarti, Shouvanik and DeCross, Matthew and Herman, Dylan and Kumar, Niraj and Larson, Jeffrey and Lykov, Danylo and Minssen, Pierre and Sun, Yue and others},
  journal={Science Advances},
  volume={10},
  number={22},
  pages={eadm6761},
  year={2024},
  publisher={American Association for the Advancement of Science}
}

@article{wang2021noise,
  title={Noise-induced barren plateaus in variational quantum algorithms},
  author={Wang, Samson and Fontana, Enrico and Cerezo, Marco and Sharma, Kunal and Sone, Akira and Cincio, Lukasz and Coles, Patrick J},
  journal={Nature communications},
  volume={12},
  number={1},
  pages={6961},
  year={2021},
  publisher={Nature Publishing Group UK London}
}

@inproceedings{swierkowska2024achieving,
  title={Achieving pareto-optimality in quantum circuit compilation via a multi-objective heuristic optimization approach},
  author={{\'S}wierkowska, Aleksandra and Echavarria, Jorge and Schulz, Laura and Schulz, Martin},
  booktitle={2024 IEEE International Conference on Quantum Computing and Engineering (QCE)},
  volume={2},
  pages={306--310},
  year={2024},
  organization={IEEE}
}

@INPROCEEDINGS{9951287,
  author={Ji, Yanjun and Brandhofer, Sebastian and Polian, Ilia},
  booktitle={2022 IEEE International Conference on Quantum Computing and Engineering (QCE)}, 
  title={Calibration-Aware Transpilation for Variational Quantum Optimization}, 
  year={2022},
  volume={},
  number={},
  pages={204-214},
  doi={10.1109/QCE53715.2022.00040}}

@article{huo2025anchor,
  title={Anchor: Reducing Temporal and Spatial Output Performance Variability on Quantum Computers},
  author={Huo, Yuqian and Leeds, Daniel and Ludmir, Jason and DiBrita, Nicholas S and Patel, Tirthak},
  journal={Proceedings of the ACM on Measurement and Analysis of Computing Systems},
  volume={9},
  number={3},
  pages={1--27},
  year={2025},
  publisher={ACM New York, NY, USA}
}

@article{cordella2004sub,
  title={A (sub) graph isomorphism algorithm for matching large graphs},
  author={Cordella, Luigi P and Foggia, Pasquale and Sansone, Carlo and Vento, Mario},
  journal={IEEE transactions on pattern analysis and machine intelligence},
  volume={26},
  number={10},
  pages={1367--1372},
  year={2004},
  publisher={IEEE}
}

@article{klimov2018fluctuations,
  title={Fluctuations of energy-relaxation times in superconducting qubits},
  author={Klimov, Paul V and Kelly, Julian and Chen, Zijun and Neeley, Matthew and Megrant, Anthony and Burkett, Brian and Barends, Rami and Arya, Kunal and Chiaro, Ben and Chen, Yu and others},
  journal={Physical review letters},
  volume={121},
  number={9},
  pages={090502},
  year={2018},
  publisher={APS}
}

@article{huang2020closer,
  title={A closer look at invalid action masking in policy gradient algorithms},
  author={Huang, Shengyi and Onta{\~n}{\'o}n, Santiago},
  journal={arXiv preprint arXiv:2006.14171},
  year={2020}
}

@article{towers2024gymnasium,
  title={Gymnasium: A standard interface for reinforcement learning environments},
  author={Towers, Mark and Kwiatkowski, Ariel and Terry, Jordan and Balis, John U and De Cola, Gianluca and Deleu, Tristan and Goul{\~a}o, Manuel and Kallinteris, Andreas and Krimmel, Markus and KG, Arjun and others},
  journal={arXiv preprint arXiv:2407.17032},
  year={2024}
}

@article{raffin2021stable,
  title={Stable-baselines3: Reliable reinforcement learning implementations},
  author={Raffin, Antonin and Hill, Ashley and Gleave, Adam and Kanervisto, Anssi and Ernestus, Maximilian and Dormann, Noah},
  journal={Journal of machine learning research},
  volume={22},
  number={268},
  pages={1--8},
  year={2021}
}

@article{schulman2017proximal,
  title={Proximal policy optimization algorithms},
  author={Schulman, John and Wolski, Filip and Dhariwal, Prafulla and Radford, Alec and Klimov, Oleg},
  journal={arXiv preprint arXiv:1707.06347},
  year={2017}
}

@article{mnih2013playing,
  title={Playing atari with deep reinforcement learning},
  author={Mnih, Volodymyr and Kavukcuoglu, Koray and Silver, David and Graves, Alex and Antonoglou, Ioannis and Wierstra, Daan and Riedmiller, Martin},
  journal={arXiv preprint arXiv:1312.5602},
  year={2013}
}

@inproceedings{mnih2016asynchronous,
  title={Asynchronous methods for deep reinforcement learning},
  author={Mnih, Volodymyr and Badia, Adria Puigdomenech and Mirza, Mehdi and Graves, Alex and Lillicrap, Timothy and Harley, Tim and Silver, David and Kavukcuoglu, Koray},
  booktitle={International conference on machine learning},
  pages={1928--1937},
  year={2016},
  organization={PmLR}
}

@article{haarnoja2018soft,
  title={Soft actor-critic algorithms and applications},
  author={Haarnoja, Tuomas and Zhou, Aurick and Hartikainen, Kristian and Tucker, George and Ha, Sehoon and Tan, Jie and Kumar, Vikash and Zhu, Henry and Gupta, Abhishek and Abbeel, Pieter and others},
  journal={arXiv preprint arXiv:1812.05905},
  year={2018}
}

@article{humble2021quantum,
  title={Quantum computers for high-performance computing},
  author={Humble, Travis S and McCaskey, Alexander and Lyakh, Dmitry I and Gowrishankar, Meenambika and Frisch, Albert and Monz, Thomas},
  journal={IEEE Micro},
  volume={41},
  number={5},
  pages={15--23},
  year={2021},
  publisher={IEEE}
}

@article{andrychowicz2020matters,
  title={What matters in on-policy reinforcement learning? a large-scale empirical study},
  author={Andrychowicz, Marcin and Raichuk, Anton and Sta{\'n}czyk, Piotr and Orsini, Manu and Girgin, Sertan and Marinier, Raphael and Hussenot, L{\'e}onard and Geist, Matthieu and Pietquin, Olivier and Michalski, Marcin and others},
  journal={arXiv preprint arXiv:2006.05990},
  year={2020}
}

@article{schulman2015high,
  title={High-dimensional continuous control using generalized advantage estimation},
  author={Schulman, John and Moritz, Philipp and Levine, Sergey and Jordan, Michael and Abbeel, Pieter},
  journal={arXiv preprint arXiv:1506.02438},
  year={2015}
}

@misc{ibmQuantumSystem,
	author = {},
	title = {{I}{B}{M} {Q}uantum {S}ystem {T}wo: the era of quantum utility is here | {I}{B}{M} {Q}uantum {C}omputing {B}log --- ibm.com},
	howpublished = {\url{https://www.ibm.com/quantum/blog/quantum-roadmap-2033}},
	year = {2023},
}

@article{chamberland2020topological,
  title={Topological and subsystem codes on low-degree graphs with flag qubits},
  author={Chamberland, Christopher and Zhu, Guanyu and Yoder, Theodore J and Hertzberg, Jared B and Cross, Andrew W},
  journal={Physical Review X},
  volume={10},
  number={1},
  pages={011022},
  year={2020},
  publisher={APS}
}

@inproceedings{wang2025accelerating,
  title={Accelerating Simulation of Quantum Circuits under Noise via Computational Reuse},
  author={Wang, Meng and Tannu, Swamit and Nair, Prashant J},
  booktitle={Proceedings of the 52nd Annual International Symposium on Computer Architecture},
  pages={1539--1553},
  year={2025}
}

@article{suzuki2021qulacs,
  title={Qulacs: a fast and versatile quantum circuit simulator for research purpose},
  author={Suzuki, Yasunari and Kawase, Yoshiaki and Masumura, Yuya and Hiraga, Yuria and Nakadai, Masahiro and Chen, Jiabao and Nakanishi, Ken M and Mitarai, Kosuke and Imai, Ryosuke and Tamiya, Shiro and others},
  journal={Quantum},
  volume={5},
  pages={559},
  year={2021},
  publisher={Verein zur F{\"o}rderung des Open Access Publizierens in den Quantenwissenschaften}
}

@article{nguyen2024qfaas,
  title={Qfaas: A serverless function-as-a-service framework for quantum computing},
  author={Nguyen, Hoa T and Usman, Muhammad and Buyya, Rajkumar},
  journal={Future Generation Computer Systems},
  volume={154},
  pages={281--300},
  year={2024},
  publisher={Elsevier}
}

@article{li2024quarl,
  title={Quarl: A learning-based quantum circuit optimizer},
  author={Li, Zikun and Peng, Jinjun and Mei, Yixuan and Lin, Sina and Wu, Yi and Padon, Oded and Jia, Zhihao},
  journal={Proceedings of the ACM on Programming Languages},
  volume={8},
  number={OOPSLA1},
  pages={555--582},
  year={2024},
  publisher={ACM New York, NY, USA}
}

@inproceedings{akiba2019optuna,
  title={Optuna: A next-generation hyperparameter optimization framework},
  author={Akiba, Takuya and Sano, Shotaro and Yanase, Toshihiko and Ohta, Takeru and Koyama, Masanori},
  booktitle={Proceedings of the 25th ACM SIGKDD international conference on knowledge discovery \& data mining},
  pages={2623--2631},
  year={2019}
}

@article{hansen2001completely,
  title={Completely derandomized self-adaptation in evolution strategies},
  author={Hansen, Nikolaus and Ostermeier, Andreas},
  journal={Evolutionary computation},
  volume={9},
  number={2},
  pages={159--195},
  year={2001},
  publisher={MIT Press}
}

@article{bharti2022noisy,
  author = {Kishor Bharti and Richard Helsen and Ryan Babbush},
  title = {Noisy Intermediate-Scale Quantum Algorithms},
  journal = {Reviews of Modern Physics},
  volume = {94},
  number = {1},
  pages = {015004},
  year = {2022}
}

@inproceedings{das2023imitation,
  title={The imitation game: Leveraging copycats for robust native gate selection in nisq programs},
  author={Das, Poulami and Kessler, Eric and Shi, Yunong},
  booktitle={2023 IEEE International Symposium on High-Performance Computer Architecture (HPCA)},
  pages={787--801},
  year={2023},
  organization={IEEE}
}

@article{gokhale2020optimized,
  title={{Optimized Quantum Compilation for Near-Term Algorithms with OpenPulse}},
  author={Gokhale, Pranav and Javadi-Abhari, Ali and Earnest, Nathan and Shi, Yunong and Chong, Frederic T},
  journal={arXiv preprint arXiv:2004.11205},
  year={2020}
}

@inproceedings{tannu2019not,
  title={{Not All Qubits are Created Equal: A Case for Variability-Aware Policies for NISQ-Era Quantum Computers}},
  author={Tannu, Swamit S and Qureshi, Moinuddin K},
  booktitle={Proceedings of the Twenty-Fourth International Conference on Architectural Support for Programming Languages and Operating Systems},
  pages={987--999},
  year={2019},
  organization={ACM}
}

@inproceedings{murali2019noise,
  title={{Noise-Adaptive Compiler Mappings for Noisy Intermediate-Scale Quantum Computers}},
  author={Murali, Prakash and Baker, Jonathan M and Javadi-Abhari, Ali and Chong, Frederic T and Martonosi, Margaret},
  booktitle={Proceedings of the Twenty-Fourth International Conference on Architectural Support for Programming Languages and Operating Systems},
  pages={1015--1029},
  year={2019},
  organization={ACM}
}

@inproceedings{wilson2021empirical,
  title={{Empirical Evaluation of Circuit Approximations on Noisy Quantum Devices}},
  author={Wilson, Ellis and Mueller, Frank and Bassman, Lindsay and Iancu, Costin},
  booktitle={Proceedings of the International Conference for High Performance Computing, Networking, Storage and Analysis},
  pages={1--15},
  year={2021}
}

@article{farhi2016quantum,
  title={{Quantum Supremacy through the Quantum Approximate Optimization Algorithm}},
  author={Farhi, Edward and Harrow, Aram W},
  journal={arXiv preprint arXiv:1602.07674},
  year={2016}
}

@article{peruzzo2014variational,
  title={A variational eigenvalue solver on a photonic quantum processor},
  author={Peruzzo, Alberto and McClean, Jarrod and Shadbolt, Peter and Yung, Man-Hong and Zhou, Xiao-Qi and Love, Peter J and Aspuru-Guzik, Al{\'a}n and O’brien, Jeremy L},
  journal={Nature communications},
  volume={5},
  number={1},
  pages={4213},
  year={2014},
  publisher={Nature Publishing Group UK London}
}

@article{cerezo2021variational,
  title={{Variational Quantum Algorithms}},
  author={Cerezo, Marco and Arrasmith, Andrew and Babbush, Ryan and Benjamin, Simon C and Endo, Suguru and Fujii, Keisuke and McClean, Jarrod R and Mitarai, Kosuke and Yuan, Xiao and Cincio, Lukasz and Coles, Patrick},
  journal={Nature Reviews Physics},
  volume={3},
  number={9},
  pages={625--644},
  year={2021},
  publisher={Nature Publishing Group UK London}
}

@misc{qiskit2024,
      title={Quantum computing with {Q}iskit},
      author={Javadi-Abhari, Ali and Treinish, Matthew and Krsulich, Kevin and Wood, Christopher J. and Lishman, Jake and Gacon, Julien and Martiel, Simon and Nation, Paul D. and Bishop, Lev S. and Cross, Andrew W. and Johnson, Blake R. and Gambetta, Jay M.},
      year={2024},
      doi={10.48550/arXiv.2405.08810},
      eprint={2405.08810},
      archivePrefix={arXiv},
      primaryClass={quant-ph}
}

@inproceedings{li2022paulihedral,
  title={{Paulihedral: A Generalized Block-wise Compiler Optimization Framework for Quantum Simulation Kernels}},
  author={Li, Gushu and Wu, Anbang and Shi, Yunong and Javadi-Abhari, Ali and Ding, Yufei and Xie, Yuan},
  booktitle={Proceedings of the 27th ACM International Conference on Architectural Support for Programming Languages and Operating Systems},
  pages={554--569},
  year={2022}
}

@article{farhi2014quantum,
  title={{A Quantum Approximate Optimization Algorithm}},
  author={Farhi, Edward and Goldstone, Jeffrey and Gutmann, Sam},
  journal={arXiv preprint arXiv:1411.4028},
  year={2014}
}

@article{havlivcek2019supervised,
  title={{Supervised Learning with Quantum-Enhanced Feature Spaces}},
  author={Havl{\'\i}{\v{c}}ek, Vojt{\v{e}}ch and C{\'o}rcoles, Antonio D and Temme, Kristan and Harrow, Aram W and Kandala, Abhinav and Chow, Jerry M and Gambetta, Jay M},
  journal={Nature},
  volume={567},
  number={7747},
  pages={209--212},
  year={2019},
  publisher={Nature Publishing Group}
}

@article{biamonte2017quantum,
  title={Quantum machine learning},
  author={Biamonte, Jacob and Wittek, Peter and Pancotti, Nicola and Rebentrost, Patrick and Wiebe, Nathan and Lloyd, Seth},
  journal={Nature},
  volume={549},
  number={7671},
  pages={195--202},
  year={2017},
  publisher={Nature Publishing Group}
}

@inproceedings{patel2022charter,
  title={Charter: Identifying the most-critical gate operations in quantum circuits via amplified gate reversibility},
  author={Patel, Tirthak and Silver, Daniel and Tiwari, Devesh},
  booktitle={SC22: International Conference for High Performance Computing, Networking, Storage and Analysis},
  pages={1--16},
  year={2022},
  organization={IEEE}
}

@inproceedings{silver2024lexiql,
  title={{LexiQL: Quantum Natural Language Processing on NISQ Machines}},
  booktitle={Proceedings of the International Conference for High-Performance Computing, Networking, Storage and Analysis},
  author={Silver, Daniel and Ranjan, Aditya and Achutha, Rakesh and Patel, Tirthak and Tiwari, Devesh},
  year={2024}
}

@article{koch2007charge,
  title={Charge-insensitive qubit design derived from the Cooper pair box},
  author={Koch, Jens and Yu, Terri M and Gambetta, Jay and Houck, Andrew A and Schuster, David I and Majer, Johannes and Blais, Alexandre and Devoret, Michel H and Girvin, Steven M and Schoelkopf, Robert J},
  journal={Physical Review A—Atomic, Molecular, and Optical Physics},
  volume={76},
  number={4},
  pages={042319},
  year={2007},
  publisher={APS}
}

@article{arute2019quantum,
  title={Quantum supremacy using a programmable superconducting processor},
  author={Arute, Frank and Arya, Kunal and Babbush, Ryan and Bacon, Dave and Bardin, Joseph C and Barends, Rami and Biswas, Rupak and Boixo, Sergio and Brandao, Fernando GSL and Buell, David A and others},
  journal={Nature},
  volume={574},
  number={7779},
  pages={505--510},
  year={2019},
  publisher={Nature Publishing Group UK London}
}

@article{preskill2018quantum,
  title={Quantum computing in the NISQ era and beyond},
  author={Preskill, John},
  journal={Quantum},
  volume={2},
  pages={79},
  year={2018},
  publisher={Verein zur F{\"o}rderung des Open Access Publizierens in den Quantenwissenschaften}
}

@article{javadi2024quantum,
  title={Quantum computing with Qiskit},
  author={Javadi-Abhari, Ali and Treinish, Matthew and Krsulich, Kevin and Wood, Christopher J and Lishman, Jake and Gacon, Julien and Martiel, Simon and Nation, Paul D and Bishop, Lev S and Cross, Andrew W and others},
  journal={arXiv preprint arXiv:2405.08810},
  year={2024}
}

@article{kim2023evidence,
  title={Evidence for the utility of quantum computing before fault tolerance},
  author={Kim, Youngseok and Eddins, Andrew and Anand, Sajant and Wei, Ken Xuan and Van Den Berg, Ewout and Rosenblatt, Sami and Nayfeh, Hasan and Wu, Yantao and Zaletel, Michael and Temme, Kristan and others},
  journal={Nature},
  volume={618},
  number={7965},
  pages={500--505},
  year={2023},
  publisher={Nature Publishing Group UK London}
}

@article{quetschlich2023mqt,
  title={MQT Bench: Benchmarking software and design automation tools for quantum computing},
  author={Quetschlich, Nils and Burgholzer, Lukas and Wille, Robert},
  journal={Quantum},
  volume={7},
  pages={1062},
  year={2023},
  publisher={Verein zur F{\"o}rderung des Open Access Publizierens in den Quantenwissenschaften}
}

@article{krantz2019quantum,
  title={A quantum engineer's guide to superconducting qubits},
  author={Krantz, Philip and Kjaergaard, Morten and Yan, Fei and Orlando, Terry P and Gustavsson, Simon and Oliver, William D},
  journal={Applied physics reviews},
  volume={6},
  number={2},
  year={2019},
  publisher={AIP Publishing}
}

@article{lubinski2023application,
  title={Application-oriented performance benchmarks for quantum computing},
  author={Lubinski, Thomas and Johri, Sonika and Varosy, Paul and Coleman, Jeremiah and Zhao, Luning and Necaise, Jason and Baldwin, Charles H and Mayer, Karl and Proctor, Timothy},
  journal={IEEE Transactions on Quantum Engineering},
  volume={4},
  pages={1--32},
  year={2023},
  publisher={IEEE}
}

@article{sivarajah2021t,
  title={t| ket>: a retargetable compiler for NISQ devices},
  author={Sivarajah, Seyon and Dilkes, Silas and Cowtan, Alexander and Simmons, Will and Edgington, Alec and Duncan, Ross},
  journal={Quantum Science \& Technology},
  volume={6},
  number={1},
  pages={014003},
  year={2021},
  publisher={IOP Publishing}
}

@inproceedings{li2019tackling,
  title={Tackling the qubit mapping problem for NISQ-era quantum devices},
  author={Li, Gushu and Ding, Yufei and Xie, Yuan},
  booktitle={Proceedings of the twenty-fourth international conference on architectural support for programming languages and operating systems},
  pages={1001--1014},
  year={2019}
}

@article{voichick2023qunity,
  title={Qunity: A unified language for quantum and classical computing},
  author={Voichick, Finn and Li, Liyi and Rand, Robert and Hicks, Michael},
  journal={Proceedings of the ACM on Programming Languages},
  volume={7},
  number={POPL},
  pages={921--951},
  year={2023},
  publisher={ACM New York, NY, USA}
}

@article{alexeev2024quantum,
  title={Quantum-centric supercomputing for materials science: A perspective on challenges and future directions},
  author={Alexeev, Yuri and Amsler, Maximilian and Barroca, Marco Antonio and Bassini, Sanzio and Battelle, Torey and Camps, Daan and Casanova, David and Choi, Young Jay and Chong, Frederic T and Chung, Charles and others},
  journal={Future Generation Computer Systems},
  volume={160},
  pages={666--710},
  year={2024},
  publisher={Elsevier}
}

@inproceedings{wilson2020just,
  title={Just-in-time quantum circuit transpilation reduces noise},
  author={Wilson, Ellis and Singh, Sudhakar and Mueller, Frank},
  booktitle={2020 IEEE international conference on quantum computing and engineering (QCE)},
  pages={345--355},
  year={2020},
  organization={IEEE}
}

@inproceedings{wang2024rate,
  title={Rate adjustable bivariate bicycle codes for quantum error correction},
  author={Wang, Ming and Mueller, Frank},
  booktitle={2024 IEEE International Conference on Quantum Computing and Engineering (QCE)},
  volume={2},
  pages={412--413},
  year={2024},
  organization={IEEE}
}

@inproceedings{chundury2025scaling,
  title={Scaling Hybrid Quantum--HPC Applications with the Quantum Framework},
  author={Chundury, Srikar and Shehata, Amir and Kim, Seongmin and Gopalakrishnan Meena, Muralikrishnan and Lu, Chao and Gottiparthi, Kalyana and Perez, Eduardo Antonio Coello and Mueller, Frank and Suh, In-Saeng},
  booktitle={Proceedings of the SC'25 Workshops of the International Conference for High Performance Computing, Networking, Storage and Analysis},
  pages={1888--1897},
  year={2025}
}

@inproceedings{mahesh2025conqure,
  title={CONQURE: A co-execution environment for quantum and classical resources},
  author={Mahesh, Atulya and Mueller, Frank},
  booktitle={2025 IEEE International Conference on Quantum Computing and Engineering (QCE)},
  volume={2},
  pages={41--45},
  year={2025},
  organization={IEEE}
}

@article{xu2025gpu,
  title={GPU-Accelerated Distributed QAOA on Large-scale HPC Ecosystems},
  author={Xu, Zhihao and Chundury, Srikar and Kim, Seongmin and Shehata, Amir and Li, Xinyi and Li, Ang and Luo, Tengfei and Mueller, Frank and Suh, In-Saeng},
  journal={arXiv preprint arXiv:2506.10531},
  year={2025}
}

@techreport{chundury2024qfw,
  title={QFw: A Quantum Framework for Large-scale HPC Ecosystems},
  author={Chundury, Srikar and Shehata, Amir and Naughton III, Thomas and Kim, Seongmin and Mueller, Frank and Suh, In-Saeng},
  year={2024},
  institution={Oak Ridge National Laboratory (ORNL), Oak Ridge, TN (United States)}
}

@inproceedings{chundury2025quantum,
  title={Quantum Simulators and Applications on Quantum Framework},
  author={Chundury, Srikar and Xu, Zhihao and Shehata, Amir and Kim, Seongmin and Mueller, Frank and Suh, In-Saeng},
  booktitle={2025 IEEE International Conference on Quantum Computing and Engineering (QCE)},
  volume={2},
  pages={522--523},
  year={2025},
  organization={IEEE}
}

@inproceedings{mittal2025openmp,
  title={OpenMP-Q: Quantum Task Offloading in OpenMP},
  author={Mittal, Swastik and Mahesh, Atulya and Mueller, Frank},
  booktitle={International Workshop on OpenMP},
  pages={81--95},
  year={2025},
  organization={Springer}
}

@inproceedings{zhang2023disq,
  title={Disq: Dynamic iteration skipping for variational quantum algorithms},
  author={Zhang, Junyao and Wang, Hanrui and Ravi, Gokul Subramanian and Chong, Frederic T and Han, Song and Mueller, Frank and Chen, Yiran},
  booktitle={2023 IEEE International Conference on Quantum Computing and Engineering (QCE)},
  volume={1},
  pages={1062--1073},
  year={2023},
  organization={IEEE}
}

\end{document}